\documentclass[aps,prx,twocolumn,10pt]{revtex4-2} %

\usepackage{amsmath} 
\usepackage{amssymb}
\usepackage{amsthm}
\usepackage{mathrsfs}
\usepackage{amsbsy}
\usepackage{bm}
\usepackage{graphicx}
\usepackage{times}
\usepackage{array}
\usepackage{upgreek}
\usepackage[dvipsnames,table]{xcolor}
\usepackage{dsfont}
\usepackage{textpos}
\usepackage{braket}
\usepackage[export]{adjustbox}\usepackage[colorlinks=true,citecolor=NavyBlue,linkcolor=RubineRed,urlcolor=Cerulean,pdfencoding=auto]{hyperref}
\usepackage{tabularray}
\UseTblrLibrary{booktabs}

\newcommand{\eref}[1]{Eq.~\eqref{#1}}
\newcommand{\fref}[1]{Fig.~\ref{#1}}
\newcommand{\hc}{^\dagger}
\newcommand{\tran}{^{\top\kern-\scriptspace}}
\newcommand\isotope[2]{\textsuperscript{#2}#1}

\DeclareMathOperator{\tr}{Tr} 
\DeclareMathOperator{\diag}{diag}

\renewcommand\labelenumi{(\arabic{enumi})}
\renewcommand\theenumi\labelenumi

\def\figwidth{\columnwidth}

\begin{document}

\title{Qubit-qudit entanglement transfer in defect centers with high-spin nuclei}

\author{Wolf-Rüdiger~Hannes}
\author{Guido~Burkard}
\affiliation{Department of Physics and IQST, University of Konstanz, 78457 Konstanz, Germany }

\begin{abstract}
	We propose a scheme for accumulating entanglement between long-lived qudits provided by central nuclear spins of defect centers. 
	Assuming a generic setting, the electron spin of each node acts as the communication qubit and may be entangled with other nodes, e.g., through a spin-photon interface. 
	The generally available Ising component of the hyperfine interaction is shown to facilitate repeated entanglement transfer onto memory qudits of arbitrary dimension $d\le 2I+1$ with $I$ the nuclear spin quantum number. 
	When $d$ is set to an integer power of two, maximal entanglement can be generated deterministically and without intermittent driving of nuclear spins.
	The scheme is applicable to several candidate systems, including the \isotope{Ge}{73} germanium vacancy in diamond.
\end{abstract}

\maketitle	

\section{Introduction}
\label{sec:introduction}

High-spin nuclei in solids comprise a new and rapidly developing  platform for quantum information processing (QIP)~\cite{FernandezdeFuentes2024}. 
In addition to their superior coherence times, such spins offer the advantage of a large Hilbert space, allowing for advanced techniques in various QIP strategies.
In the field of quantum error correction, their spin subspace has been identified as a
hardware-efficient structure for encoding fault-tolerant logical qubits~\cite{Gross2021a,Lim2023,Lim2025}.
Advantages over traditional spin qubits could also arise in QIP strategies which do not involve quantum error correction on the lowest level, but instead use the multilevel space directly as a qudit.
This includes both the field of quantum communication (QC)~\cite{Nemoto2016} and measurement-based quantum computing (MBQC)~\cite{Briegel2001,Raussendorf2001,Raussendorf2006}.
Several theoretical studies~\cite{Zhou2003,Tame2006,Wang2017a} have addressed high-dimensional one-way quantum processing, while experimentally, first proof-of-concept operations have been demonstrated~\cite{Reimer2019}.

A promising route towards scalable quantum networking is based on devices comprising several modules, where each module consists of an optically addressable defect center in a solid matrix, featuring both an electron and one (or more) nuclear spins, which serve as communication and memory qubits, respectively. 
In this context, a single network node~\cite{Stas2022} and a two-node quantum network with remote entanglement storage~\cite{Knaut2024} have recently been realized, both works based on silicon-vacancy centers (SiVs) in diamond-based nanophotonic cavities.
At the same time, a higher-dimensional nuclear-spin platform with very similar properties is currently being established~\cite{Adambukulam2024}, 
namely the \isotope{Ge}{73} germanium vacancy (GeV) in diamond with nuclear spin $I=9/2$.
Thus, it is foreseeable that at some point such a quantum network will be extended to high-spin nuclei with the above-mentioned advantages. 
Another promising type of higher-dimensional nuclear-spin platforms is provided by implanted group V donors in silicon. 
In this case, the route towards scaling up is somewhat different from the above, and several methods of coupling over microscopic distances have been proposed~\cite{Morello2020}. However, most of these methods, e.g., the coupling via microwave photons, connect the nuclear spin qudits indirectly via their respective electron spin qubits so that the situation is in principle similar to the optical network device with macroscopically separate nodes. 

A natural question for such systems, in particular in view of higher-dimensional MBQC and QC protocols, is whether 
a maximally entangled state can be generated between two distant memory qudits, if they only interact with their respective electron spin qubit. Here we provide a solution to this problem considering a hyperfine tensor structure common for central nuclear spins. Enlarging the nuclear qubit memory based on the \isotope{N}{15} nitrogen vacancy (NV) center~\cite{Nemoto2014} to qudit dimesions $d>2$, we show that the Ising part may be exploited to realize repeated entanglement transfer, also known as entanglement accumulation~\cite{Giordani2021}. 
A driving-free and particularly simple version of an iterative scheme results in maximal entanglement deterministically if ${d}$ is an integer power of two, or with success rate $1/{d}$ for all other values of~${d}$. 
For specific unfavorable qudit dimensions~${d}$, other schemes without driving and with higher transfer success rates are constructed, and the inclusion of nuclear-spin driving in between iterations is shown to enable deterministic protocols. 
Another option pursued for unfavorable dimensions~${d}$ is the generation of highly but partially entangled states without driving. Such states may still be useful to some extent, although they do not allow perfectly reliable QC protocols~\cite{Delgado2005}. 

In the case of the \isotope{NV}{15} center, driving the nuclear spin is slow due to the small gyromagnetic ratio, but can be enhanced by the hyperfine coupling~\cite{Nemoto2014}. For higher spins, while the hyperfine coupling is typically much stronger, such operations could still be relatively slow due to the large number of nearly resonant transitions. 
Thus, we mainly focus on schemes where driving is required only for 
initialization and, if desired, final state correction. Apart from possible differences in single-qubit electron gate implementation, these schemes do not pose any additional demands on the controlling hardware in comparison with the qubit scheme~\cite{Nemoto2014}.
In the bipartite scenario, our focus is on generating \textit{any} maximally entangled state, and we discuss the explicit form of the state only in special cases. 

The model on which our scheme is based is applicable to a wide range of nuclear spins $I\ge 1$ in appropriate parameter regimes. 
In diamond, aside from the GeV~\cite{Adambukulam2024} there is also the \isotope{NV}{14} center with $I=1$ which could allow qutrit applications~\cite{Fu2022}. 
Considering other host materials, the vanadium \isotope{V}{51} defect in silicon carbide~\cite{Koller2025} with $I=7/2$ is recently attracting attention.
A promising category, featuring very strong electron–nuclear coupling,  are implanted group-V donors in silicon, 
in particular 
\isotope{As}{75} with $I=3/2$~\cite{Franke2015},
\isotope{Sb}{123} with $I=7/2$~\cite{Asaad2020,FernandezdeFuentes2024,Yu2025},
and \isotope{Bi}{209} with $I=9/2$~\cite{Morley2010}.
For all the above defects, the electron-nuclear spin dynamics is explicitly shown to qualify, with larger imperfections only in one case, the V defect.
Another emerging platform consists of single rare-earth ions, some of which can be made optically addressable by single photons~\cite{Zhong2018,Dibos2018} and strongly couple to (e.g., aluminium nuclear spins with $I=5/2$~\cite{Siyushev2014}) or host (e.g., \isotope{Pr}{141} with $I=5/2$~\cite{Nakamura2014,Utikal2014}, \isotope{Nd}{143} with $I=7/2$, or \isotope{Er}{167} with $I=7/2$) high nuclear spins, even though coherent control would have to be demonstrated for some of those examples. 
While two of these spin quantum numbers (3/2, 7/2) just fit a 
power-of-two dimensional qudit, 
applying the driving-free scheme to others is possible by lowering $d$ from $2I+1$ to the next favorable number, which can be achieved by leaving the difference number of levels unoccupied during initialization. 

The paper is structured as follows. 
In Sec.~\ref{sec:methods} we describe the problem of qubit-qudit entanglement transfer in a general context and derive explicit conditions to be fulfilled in specific iterations.
Sec.~\ref{sec:network_model} contains the model for a single network node and a description of the photonic scheme as an example mechanism for generating internode entanglement links. 
Using this model, entanglement accumulation schemes are constructed for two nodes in Sec.~\ref{sec:schemes}, including both deterministic and probabilistic protocols. 
The successive application of one selected scheme to pairs of nodes in a larger quantum network is demonstrated to result in higher-dimensional multipartite entanglement in Sec.~\ref{sec:multipartite}.
Some further technical aspects are discussed in Sec.~\ref{sec:discussion}. 

\section{Basic principles}
\label{sec:methods}

\subsection{Entanglement measure and effective gate}
\label{sec:theory_entanglement}

\begin{figure}[t]
	\centering
	\includegraphics[width=\figwidth]{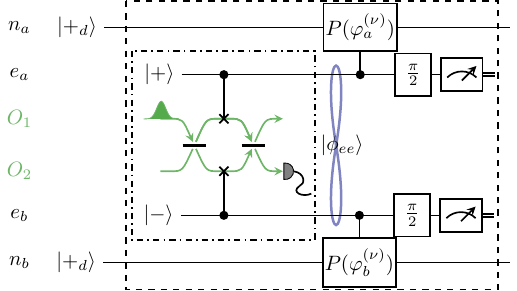}
	\caption{Scheme for iteratively entangling two memory qudits $n_a,n_b$. The dash-dotted box contains the photonic setup with optical arms $O_1$ and $O_2$ for entangling remote electron spin qubits $e_a,e_b$, producing a state $\ket{\phi_{ee}}$, usually an EPR pair. The dashed box corresponds to one iteration of entanglement transfer. 
	The $\frac{\pi}{2}$-gates are $Y^{1/2}$ rotations. The equal superposition state $\ket{+_d}$ is defined in \eqref{eq:psi0}.
	}
	\label{fig:circuit}
\end{figure}

We consider two nodes labeled by the index ${K}=a,b$, each consisting of an ancillary qubit (states $\ket{0}$ and $\ket{1}$) and a memory qudit of dimension~$d$. 
While this section is model-independent, we refer to the qubit and qudit as electron and nuclear spin, respectively.
Prior to the scheme, nuclear spins are initialized in a suitable product state $\ket{\psi^{(0)}}$. 
One iteration of entanglement transfer consists of three steps: (i) generation of an entangled state (denoted $\ket{\phi_{ee}}$) of the two distant electron spins,
(ii) in each node ${K}$ a controlled gate, denoted $\text{C}(U_{K,0},U_{K,1})=\ket{0}\bra{0} \otimes U_{K,0} + \ket{1}\bra{1} \otimes U_{K,1}$, between electron and nuclear spins, where $U_{K,0}$ and $U_{K,1}$ commute,
and
(iii) measurement of both electron spins in the $X$ basis (or along any other axis perpendicular to $Z$), with outcomes denoted by $j_a, j_b\ \in\{0,1\}$.
The state into which nuclear spins are projected depends on the measurement outcomes, and its entanglement may generally also depend. 
We call the process a deterministic complete entanglement transfer if for any measurement outcome, the entanglement is fully transferred. 
\fref{fig:circuit} shows a circuit with $\text{C}(U_{K,0},U_{K,1})$ realized by CPHASE gates, fitting to the model to be considered later. 

First, we introduce a method to assess two-qudit entanglement. 
Consider some bases $\{\ket{u_i}_a\}$ and $\{\ket{u_i}_b\}$ for the two subsystems, 
where $i=1,\dots, d$. %
A general pure two-qudit state is expanded in this basis as $\ket{\psi} = \sum_{{i_a},{i_b}}^{} c_{i_a,i_b} {\ket{u_{i_a}}_a \otimes \ket{u_{i_b}}_b}$, with the norm $\sum_{{i_a},{i_b}}^{} |c_{{i_a},{i_b}}|^2$. 
It is convenient to define a ${d}\times {d}$ matrix ${{\psi}}$ from the coefficients so that the vectorization of ${{\psi}}\tran$ gives the state vector $\ket{\psi}$ in this basis.
Local operations, $O_a \otimes O_b \ket{\psi}$, are translated as~$O_a \psi O_b\tran$. The Schmidt decomposition of $\ket{\psi}$ corresponds to the singular value decomposition of ${{\psi}}$, with the singular values or Schmidt coefficients denoted by $\{\sqrt{\chi_k}\}$ and fulfilling $\sum_k \chi_k = 1$. 
The reduced density matrix $\rho_a = \tr_b \ket{\psi} \bra{\psi}$ is given by $\rho_a = {{\psi}} {{\psi}}\hc$ (and $\rho_b = ({{\psi}}\hc {{\psi}})\tran$) and its eigenvalues are $\{\chi_k\}$.
For pure states, a unique entanglement measure exists, given by the 
von Neumann entropy of the reduced density matrix ($\rho_a$ or $\rho_b$),
or equivalently the Shannon entropy of the squared Schmidt coefficients,
\begin{align}
	\label{eq:entanglementmeasure}
	E %
	&= - \tr \rho_a \log_2\!\rho_a 
	= - \sum_k \chi_k \log_2\chi_k.
\end{align}
The state $\ket{\psi}$ is maximally entangled with $E=\log_2{d}\equiv E_{d}$ if and only if all Schmidt coefficients are equal, $\chi_k = 1/d$, and hence, if and only if the matrix $\sqrt{d}\psi$ is unitary. The number of non-zero Schmidt coefficients is denoted as the Schmidt rank~$r=\mathrm{rank}({{{\psi}}})$ and may be interpreted as the dimensionality of entanglement \cite{Friis2019}.

For later use, we briefly apply this measure to a general two-qubit state (of electron spins), expanded in the computational basis as 
\begin{equation}
	\label{eq:2qubitstate}
	\ket{\phi_{ee}} = \sum_{j_a=0,1}\sum_{j_b=0,1}^{} c_{j_a j_b} \ket{j_a, j_b}.
\end{equation}
The two Schmidt coefficients are $\sqrt{\lambda_{\pm}}$ with
\begin{equation}
	\label{eq:Schmidt_ee}
	\lambda_{\pm} = \frac{1}{2} \pm \sqrt{\frac{1}{4}-(\det \phi_{ee})^2}, \quad
	\phi_{ee} = \begin{pmatrix}
		c_{00} & c_{01} \\ c_{10} & c_{11}
	\end{pmatrix}.
\end{equation}
The entanglement is $E_{ee} = - \lambda_+ \log_2 \lambda_+ - \lambda_- \log_2 \lambda_-$.
Maximal entanglement $E_{ee}=1$ is obtained for $|\det \phi_{ee}| = 1/2$.

Next we derive an effective action in the nuclear spin space describing one iteration of entanglement transfer,
${\ket{\psi^{(\nu)}_{j_a j_b}} = {\cal T}_{j_a j_b} \ket{\psi^{(\nu-1)}}}$,
where $\ket{\psi^{(\nu)}}$ is the nuclear state after iteration~$\nu$. Here, the two indices denote the measurement outcomes ($j_a,j_b$) in round $\nu$, while (for $\nu>1$) the dependence of the states on the previous measurement outcomes is taken as implicit.
We assume that the internode entangling mechanism, described in Sec.~\ref{sec:el-el-entangling}, leaves nuclear spins decoupled until successful generation of an entangled state $\ket{\phi_{ee}}$, so here we only need to include the electron-nuclear controlled gate and the $X$ basis measurement of electron spins. 
In the basis $\{e_a\}\otimes \{e_b\}\otimes \{n_a\}\otimes \{n_b\}$, where, e.g. $\{e_a\}=(\ket{0},\ket{1})$ is the basis of the electron spin qubit in node $a$,
the controlled operation ${\cal M} = \text{C}(U_{a,0},U_{a,1}) \otimes \text{C}(U_{b,0},U_{b,1})$
takes the block-diagonal form,
\begin{align}
	{\cal M} &= \begin{pmatrix}
		U_{a,0} \otimes U_{b,0} & & & \\ & U_{a,0} \otimes U_{b,1} & & \\ & & U_{a,1} \otimes U_{b,0} & \\ & & & U_{a,1} \otimes U_{b,1}
	\end{pmatrix},
\end{align}
with all off-diagonal blocks being zero. 
To implement the $X$~basis measurement we further apply the electron spin $\pi/2$~rotations, resulting in
$(Y^{1/2} \otimes Y^{1/2}) {\cal M}$.
The effective gate ${\cal T}_{j_a j_b}$ is obtained by projecting onto the measured~($\ket{j_a,j_b}$) and initial state ($\ket{\phi_{ee}}$, prepared, e.g., by a successful photonic scheme attempt) of both electron spins from the left and right, respectively, 
\begin{equation}
	{\cal T}_{j_a j_b} = \bra{j_a,j_b} (Y^{1/2} \otimes Y^{1/2}){\cal M} \ket{\phi_{ee}}.
\end{equation}
For a general two-qubit state, \eref{eq:2qubitstate}, we explicitly obtain
\begin{equation}
	\label{eq:Tmat}
	{\cal T}_{j_a j_b} = \frac{1}{2} \sum_{j_a'j_b'} \eta_{j_a j_a'} \eta_{j_b j_b'} c_{j_a'j_b'} U_{a,j_a'}\otimes U_{b,j_b'},
\end{equation}
where $\eta_{j j'}\equiv (-1)^{\delta_{j,0}\delta_{j',1}}$.
The effective gate ${\cal T}_{j_a j_b}$ is in general not unitary. 
Without normalization, the state resulting from the entanglement transfer is, 
in matrix notation,
\begin{equation}
	\label{eq:psinu}
	\psi^{(\nu)}_{j_a j_b} = \frac{1}{2} \sum_{j_a'j_b'} \eta_{j_a j_a'} \eta_{j_b j_b'} c_{j_a'j_b'} U_{a,j_a'} \psi^{(\nu-1)} U_{b,j_b'}\tran.
\end{equation}
Note that the subscripts $j_a, j_b$ are never meant as matrix element indices in nuclear spin space.
The probability for obtaining the measurement outcome $(j_a,j_b)$ is given by
$P^{(\nu)}_{j_a j_b}=\braket{\psi^{(\nu)}_{j_a j_b}|\psi^{(\nu)}_{j_a j_b}}=\tr \psi^{(\nu)}_{j_a j_b} \psi^{(\nu)\dagger}_{j_a j_b}$ provided $\ket{\psi^{(\nu-1)}}$ has been normalized.
The entanglement $E^{(\nu)}_{j_a j_b}$ generally also depends on the measurement outcome; when we omit the measurement outcome indices, we mean the target value $E^{(\nu)}$ that we want to achieve. 

\begin{figure}[t]
	\centering
	\includegraphics[width=\columnwidth]{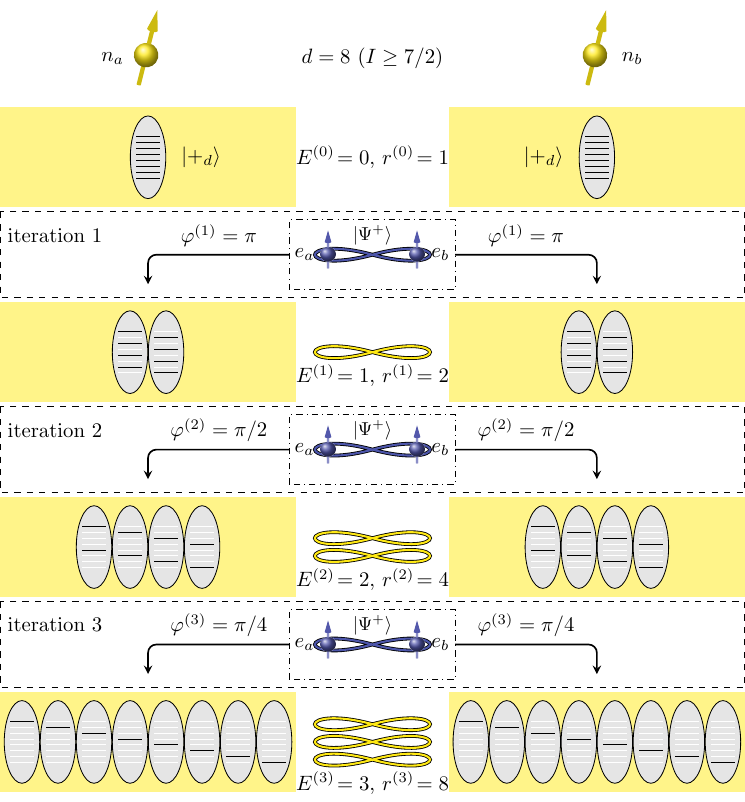}
	\caption{Illustration of a series of 1-ebit transfers for maximally entangling qudits with $d=8$ without nuclear-spin driving. The dashed and dash-dotted boxes have the same meaning as in \fref{fig:circuit}. The arrows represent entanglement transfers. Each oval shape represents a Schmidt vector, with the contained bars in black indicating the occupied levels; the initial state is $\ket{+_{d}}\otimes\ket{+_{d}}$. For this particular phase set $\{\varphi^{(\nu)}\}$, resource state $\ket{\phi_{ee}}=\ket{\Psi^{+}}$, and postselection of $j_a=j_b$, the $r^{(\nu)}$ Schmidt vectors $\{\ket{u_{k}}_K\}$ (omitting the iteration index for ease of notation) take a relatively simple form
	$\{\sqrt{r/d}\sum_{i=0}^{d/r-1} \sigma_{i,r,K}\ket{k+r i}\}_{k=0}^{r-1}$, where $\sigma_{i,r,K}$ are sign factors. The final state $\ket{\psi^{(3)}}$ is a two-qudit Bell state. 
	In the case $I>7/2$, a number of $2I+1-8$ outermost levels are unoccupied and not shown.  
	}
	\label{fig:scheme_Schmidt_8d}
\end{figure}

To illustrate the principle of entanglement accumulation we show in \fref{fig:scheme_Schmidt_8d} three iterations of a particularly simple protocol. Here the repetitive complete transfer of 1~ebit, based on suitable input choices to be derived below, implies that in each iteration the Schmidt rank doubles, $r^{(\nu+1)}=2r^{(\nu)}$, and the $r^{(\nu)}$ Schmidt coefficients of $\ket{\psi^{(\nu)}}$ are equal to $1/\sqrt{r^{(\nu)}}$. The Schmidt vectors depend on the measurement outcomes, and in the shown case ($j_a^{(\nu)}=j_b^{(\nu)}$ for all $\nu$) take a particularly simple form. 
Because the chosen phase set $\{\varphi^{(\nu)}\}$ results in a deterministic protocol with respect to entanglement generation (see Sec.~\ref{sec:scheme_deterministic}), the Schmidt coefficients are the same for other measurement outcomes, while the Schmidt vectors are related to the ones shown here by local unitaries.

\subsection{Transferability conditions}
\label{sec:transferability}

We now derive explicit conditions for achieving the goal of generating maximally entangled states.
The term `transferability' is adopted from Ref.~\cite{Giordani2021}, where conditions have been formulated for deterministic transfer of $E_{ee}=1$ to a target qudit pair, irrespective of their Hilbert space dimension. 
Their condition is formulated in terms of the Schmidt decomposition of the state immediately before the projection(s), which in our case can be written as
$\sum_{j_a j_b} \ket{j_a,j_b} {\cal T}_{j_a j_b} \ket{\psi^{(\nu-1)}}$, since the different projected electron states form an orthonormal set.
In the least explicit form, the transferability condition states that the Schmidt coefficients are to be preserved by the projection.
In this work, we take a slightly different but not necessarily less general approach by formulating conditions for entanglement transfer based on the three-step process described in Sec.~\ref{sec:theory_entanglement}. Thus, the conditions depend on resource state $\ket{\phi_{ee}}$, conditional operations $\{\text{C}(U_{K,0},U_{K,1})\}$, and, in the case of postselection, measurement outcomes $\{j_K\}$. 
The fact that $\{U_{K,0},U_{K,1}\}$ are unitary operations allows us to formulate more explicit conditions than given in Ref.~\cite{Giordani2021}. Another difference is that we consider entanglement accumulation in finite-dimensional systems, which are to be maximally entangled, and this requires different strategies, at least for some values of $d$.
We further note that the increase of the two-qudit entanglement in one iteration is bound by the consumed resource $E_{ee}$ even in the probabilistic case of postselection, despite the involvement of projections. 

To derive the following conditions, we write the state resulting from the previous iteration $\nu-1$, which for $\nu=1$ is the initial state, in the Schmidt decomposition
\begin{equation}
	\label{eq:Schmidt}
	\ket{\psi^{(\nu-1)}} = \sum_{k=1}^{r^{(\nu-1)}} \sqrt{\chi_k} \ket{u_{k}}_a \otimes \ket{u_{k}}_b.
\end{equation} 
In matrix notation, this reads
\begin{equation}
	\label{eq:Schmidt_VK}
	{\psi^{(\nu-1)}} = V_a \diag(\{\sqrt{\chi_k}\}) V_b\tran,
\end{equation} 
with semi-unitary $d \times r^{(\nu-1)}$-dimensional matrices $V_K$, whose columns correspond to $\{\ket{u_{k}}_K\}$.
The Schmidt coefficients are positive and sorted in decreasing order. 
The entanglement is $E^{(\nu-1)} = - \sum_k \chi_k \log_2\chi_k$.
Substituting into \eref{eq:psinu}, one finds that, to increase the rank by  $\delta r^{(\nu)} \equiv r^{(\nu)}-r^{(\nu-1)} \ge 0$, it is necessary (and sufficient for $E_{ee}>0$) that $\delta r^{(\nu)}$ columns of $U_{K,1} V_K$ lie outside the span (column space) of $U_{K,0} V_K$ for $K=a,b$; further one can see that the rank at most doubles, $\delta r^{(\nu)} \le r^{(\nu-1)}$.
Let us define the orthonormal basis ${Q}_{K}$ of the subspace spanned by the columns of $U_{K,0} V_K$ and $U_{K,1} V_K$.
The first $r^{(\nu-1)}$ columns of ${Q}_{K}$ are chosen as $U_{K,0} V_K$, the last $\delta r^{(\nu)}$ columns denoted as $V_K^\perp$. This allows us to write
\begin{equation}
	\label{eq:Schmidt_QK}
	U_{a,0} {\psi^{(\nu-1)}} U_{b,0}\tran  = Q_a \sqrt{S_\downarrow} Q_b\tran, 
\end{equation} 
where $\sqrt{S_\downarrow}=\diag(\{\sqrt{\chi_k}\})\oplus 0_{\delta r^{(\nu)}}$ is the direct sum of the diagonal matrix of the singular values with a zero matrix of size $\delta r^{(\nu)}$; the down arrow indicates the decreasing order of the diagonal elements. 
According to \eref{eq:psinu}, the column (row) space of $\psi^{(\nu)}_{j_a j_b}$ lies in the span of ${Q}_a$ (${Q}_b$), so we can write
\begin{equation}
	\psi^{(\nu)}_{j_a j_b} = {Q}_a \tilde{\psi}^{(\nu)}_{j_a j_b} {Q}_b\tran
\end{equation}
with $\tilde{\psi}^{(\nu)}_{j_a j_b} \equiv {Q}_{a}\hc \psi^{(\nu)}_{j_a j_b} {Q}_{b}^* $. 
Explicitly, we obtain
\begin{equation}
	\label{eq:psinu_S_sum}
	\psi^{(\nu)}_{j_a j_b} = \frac{1}{2} \sum_{j_a'j_b'} \eta_{j_a j_a'} \eta_{j_b j_b'} c_{j_a'j_b'} {\cal U}_{a,j_a'} \sqrt{S_\downarrow}  {\cal U}_{b,j_b'}\tran
\end{equation}
with ${\cal U}_{K,j} \equiv {Q}_{K}\hc {U}_{K,j} {U}_{K,0}\hc {Q}_{K}$.
Now, since ${\cal U}_{K,0} = \mathds{1}_{r^{(\nu)}}$, we define ${\cal U}_{K} \equiv {\cal U}_{K,1}$ and write out the sum as
\begin{align}
	\label{eq:psinu_S}
	\tilde{\psi}^{(\nu)}_{j_a j_b} 
	&=
	\frac{1}{2}\left(
	c_{00} \sqrt{S_\downarrow} -
	(-1)^{j_b} c_{01} \sqrt{S_\downarrow} {\cal U}_{b}\tran 
	\right. \\ & \  \left.	
	-(-1)^{j_a} c_{10} {\cal U}_{a} \sqrt{S_\downarrow} +
	(-1)^{j_a+j_b} c_{11} {\cal U}_{a} \sqrt{S_\downarrow} {\cal U}_{b}\tran \right). \nonumber
\end{align}
The matrix ${\cal U}_{K}$ is unitary in general only for $r^{(\nu)} = d$.

\subsubsection{Deterministic complete entanglement transfer}
\label{sec:condition_transfer_deterministic}

The necessary and sufficient condition for the complete deterministic transfer of $E_{ee}$ is that in each node the two gates $U_{K,j}$ generate out of the previous Schmidt vectors (the columns of $V_K$) two mutually orthogonal sets of $r^{(\nu-1)}$ vectors 
(which implies $\delta r^{(\nu)} = r^{(\nu-1)}$, thus $r^{(\nu)}$ even), which can be expressed as
\begin{align}
	\label{eq:cond_complete}
	V_K\hc {U}_{K,0}\hc U_{K,1} V_K &= 0 ,
\end{align}
for $K=a,b$.
An equivalent formulation is
\begin{equation}
	\label{eq:cond_complete_calU}
	S_\downarrow {\cal U}_K S_\downarrow = 0.
\end{equation}
To prove both directions, we note that in this case only can we choose $V_K^\perp = U_{K,1} V_K$. 
Then, ${\cal U}_{K}$ is block-antidiagonal with the lower left block given by the identity matrix $\mathds{1}_{r^{(\nu-1)}}$. 
The state from \eref{eq:psinu_S} becomes
\begin{align}
	\label{eq:psinu_complete}
	\tilde{\psi}^{(\nu)}_{j_a j_b} 
	&=
	\frac{1}{2}
	\left[\begin{pmatrix}
		1 & 0 \\ 0 & -(-1)^{j_a}
	\end{pmatrix}
	\phi_{ee}
	\begin{pmatrix}
		1 & 0 \\ 0 & -(-1)^{j_b}
	\end{pmatrix}\right]
	\nonumber\\
	&\qquad\qquad\qquad\qquad\qquad\quad
	\otimes
	\diag(\{\sqrt{\chi_k}\}).
\end{align}
The matrices with sign factors are just basis vector inversions and do not affect the singular values. 
The tensor product structure shows that the Schmidt coefficients of $\ket{\psi^{(\nu)}}$ are given by pairwise products of that of $\ket{\phi_{ee}}$ and $\ket{\psi^{(\nu-1)}}$, i.e., each~$r^{(\nu-1)}$ Schmidt coefficients $\{\sqrt{\lambda_+ \chi_k}\}$ and $\{\sqrt{\lambda_- \chi_k}\}$ with $\lambda_{\pm}$ given in \eref{eq:Schmidt_ee}. 
Straightforward algebra then shows that the entanglement, defined in \eref{eq:entanglementmeasure}, is given by
${E^{(\nu)} = E^{(\nu-1)} + E_{ee}}$.

This type of entanglement transfer has the following properties:
(i) It works independently of the choice of entanglement resource state $\ket{\phi_{ee}}$ (with $E_{ee}>0$) and also comes without any explicit constraints on the values of the $r^{(\nu-1)}$ Schmidt coefficients, apart from $r^{(\nu-1)}\le \lfloor d/2\rfloor$.
(ii)~The norm of the state $\ket{\psi^{(\nu)}_{j_a j_b}}$ is $1/2$, which means that the probabilities of individual measurement outcomes are equal, $P^{(\nu)}_{j_a j_b}=1/4$.
(iii) If during the conditional operation $\text{C}(U_{K,0},U_{K,1})$ a single-qubit $Z$ gate is applied on the electron spin $e_K$, this amounts to multiplying the Schmidt vectors of $\ket{\psi^{(\nu)}_{j_a j_b}}$ with phases, which does not affect the complete transfer. 

Since \eref{eq:cond_complete} is a necessary condition but requires $r^{(\nu)}$ to be even, it is impossible to generate an odd Schmidt rank~$r^{(\nu)}$ by a complete deterministic transfer. This includes the generation of an odd-ranked maximally entangled state. 
It might still be possible to achieve the latter goal by an incomplete deterministic transfer; the corresponding conditions are given in the next subsection (Sec.~\ref{sec:condition_Emax_deterministic}). 
Otherwise, only probabilistic entanglement transfer remains as an option to generate odd-ranked maximal entanglement.

\begin{figure}[t]
	\centering
	\includegraphics[width=\columnwidth]{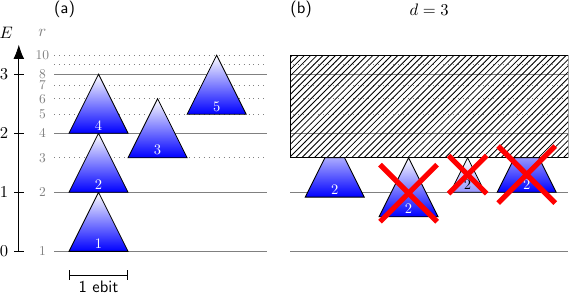}
	\caption{Some relevant cases of allowed and forbidden (crossed out) deterministic entanglement transfers. 
	Each (truncated) triangle represents an (in-)complete entanglement transfer, where the electronic entanglement $E_{ee}$ is indicated by the base width, in most cases 1~ebit, and the nuclear spin entanglement $E$ corresponds to the vertical axis. The number above the triangle base denotes the rank prior to the transfer. 
	In~(a), the target space dimension could be any number $d\ge r$ with $r$ the final rank to be reached, while in (b) the case $d=3$ is shown.
	}
	\label{fig:transferability}
\end{figure}

\subsubsection{Deterministic generation of maximal entanglement}
\label{sec:condition_Emax_deterministic}

Consider now the final iteration in which a maximally entangled state, $E^{(\nu)}=E_d$, is to be generated.
If $E^{(\nu-1)}\!=E_d-1$, then, apart from the need $E_{ee}^{(\nu)}=1$, the situation is equivalent with that considered in Sec.~\ref{sec:condition_transfer_deterministic}, and the necessary and sufficient condition is \eref{eq:cond_complete}.
Here we thus consider $E^{(\nu-1)}\!>E_d-1$, and hence, $r^{(\nu-1)}>{d}/2$. %
In Appendix~\ref{sec:criteria_derivation_max} we derive the following items as necessary and sufficient conditions:
\begin{enumerate}
	\item $\ket{\phi_{ee}^{(\nu)}}$ is maximally entangled, i.e., $E_{ee}^{(\nu)}=1$. (Thus, the entanglement transfer is incomplete.)
	\item The diagonal elements of $S_\downarrow$, consisting of squared Schmidt coefficients and zeros, come in pairs $1/{d} \pm \varepsilon_i$ with $0\le \varepsilon_i \le 1/{d}$, for $i=1,\dots,\lfloor {d}/2 \rfloor$, and one further unpaired central element $1/{d}$ 
	in the case of odd ${d}$. 
	Groups of identical successive diagonal elements of $S_\downarrow$ are termed blocks.
	\item The unitary matrix ${\cal U}_{K}$ (for $K=a,b$) has a corresponding block-antidiagonal structure. This yields an explicit condition for the physical interaction $U_K$, since the transformation $Q_K$ relating the two matrices is for $r^{(\nu)} = d$ uniquely determined, apart from trivial operations (phases/reordering) within subspace $V_K^\perp$.
	\item Two further constraints on $\{{\cal U}_{K}\}$ described by Eqs.~\eqref{eq:cond_gen3}, which can be further simplified depending on $r^{(\nu-1)}$ and the choice of $\ket{\phi_{ee}^{(\nu)}}$.
\end{enumerate}

Note that constraint (2) implies that $\delta r^{(\nu)}$ Schmidt coefficients must be equal to $2/d$, and further, that it is impossible to deterministically generate maximal entanglement for odd $d$ when the previous state is even-ranked with equal Schmidt coefficients. Therefore, a strategy derived from these constraints, which covers only the final iteration, cannot be combined with a scheme for the first iterations derived from Sec.~\ref{sec:condition_transfer_deterministic}, to give a complete scheme for odd $d$.

\subsubsection{Implications}

Before applying the transferability conditions to an actual physical model, it is worth discussing their general implications in more detail. 
The condition \eqref{eq:cond_complete} for complete transfer is particularly relevant for the first iterations, where the target Hilbert space is still sufficient to adapt each $E_{ee}^{(\nu)}$. 
This condition depends only on the Schmidt vectors obtained from the previous iteration and the electron-nuclear interaction $U_{K,j}$. If it can be repeatedly fulfilled, it allows us to maximally entangle $d=2^n$ qudits by $n$ successive transfers of 1 ebit. 
This is illustrated in \fref{fig:transferability}~(a),  along with two further example transfers starting with an odd rank. 
These are essentially all possible complete transfers reaching ranks up to ten. 
Note that because \eqref{eq:cond_complete} is independent of the resource state and inherited Schmidt coefficients, each triangle in part (a) could be made smaller (keeping the base level fixed) and/or shifted downwards towards $E=0$. However, the most relevant application of this condition is indeed the realization of successive 1-ebit transfers, where Schmidt coefficients remain equal (so that $E=\log_2 r$).

A different situation occurs in the potentially final iteration, where only limited target Hilbert space is available, as shown for $d=3$ in \fref{fig:transferability}~(b). 
The case of odd $d$ is more interesting here as it forbids a complete transfer, even regardless of $E_{ee}$ (first two crossed-out triangles).
Rather counterintuitively, this situation forces us to use an EPR pair (i.e., 1~ebit) as the resource.
In the case $d=3$, the two Schmidt coefficients obtained from the previous iteration ($\nu=1$, reaching $r^{(1)}=2$) are precisely determined as $\{\sqrt{2/3},\sqrt{1/3}\}$, and so is the entanglement $E^{(1)}=-\frac{2}{3}+\log_2 3 \approx 0.918$ (base level of the truncated triangle on the left).
In particular, as already mentioned above, for any $r \neq 2^n$ we cannot reach $E=\log_2 r$ by one rank-increasing transfer from $E=\lfloor\log_2 r\rfloor$ (third crossed-out triangle). 
The only alternative to the allowed solution shown for $d=3$, which we have not ruled out completely, is a scheme with additional iterations, some of which are rank-preserving; however, such a scheme is not expected to exist in any typical case because more iterations only lead to more constraints.

Here, we have discussed transferability conditions for deterministic entanglement accumulation. 
In principle, it is possible to derive conditions for probabilistic generation of maximal entanglement
(i.e., using postselection of electron measurement results). However, these constraints are less transparent and omitted for simplicity. 
Note that in the case of postselection, a complete transfer may be realized even when the operations $\text{C}(U_{K,0},U_{K,1})$ do not have full entangling capacity, as opposed to condition~\eqref{eq:cond_complete};
this fact is closely related with entanglement concentration through probabilistic entanglement swapping proposed in Ref.~\cite{Bose1999}.

\section{Model of the quantum network}
\label{sec:network_model}

The concept of entanglement transfer is known to be applicable to both defect centers with central nuclear spin~\cite{Nemoto2014} and with weakly coupled nuclear spins~\cite{Hannes2024}. 
In order to apply the concept in the version generalized to higher-dimensional memory qudits, we define an explicit physical model of a quantum network node as a defect center with higher nuclear spin in Sec.~\ref{sec:node_model}. A known mechanism for entangling two such nodes is described in Sec.~\ref{sec:el-el-entangling}.
In Sec.~\ref{sec:decoupling} we review the methods for using electron and nuclear spins as separate information carriers in each hybrid node, and discuss the errors accumulating on nuclear spins during each electron-electron-entangling scheme.

\subsection{Single node: Defect center}
\label{sec:node_model}

A defect center with one central electron and one nuclear spin serves as a node, and we assume that no further strongly coupled spins are nearby. 
Decoherence is not included as we focus on the basic principle; this is also to a good degree justified from the time scales as discussed in Sec.~\ref{sec:decoupling} and \ref{sec:discussion}.
We further assume that a magnetic field ${B}$ is applied along the symmetry axis of the defect, which is taken as the $z$ axis. 
The Hamiltonian of the node is composed of electron, nuclear, and hyperfine parts,
\begin{equation}
	\label{eq:H_tot}
	H = H_{e} + H_{n} + H_{\rm{hf}}.
\end{equation}

The electron part $H_{e}$ may contain several contributions depending on the type of defect center and crystal. 
In this Section, we describe the simplest case without additional 
contributions such as strain and orbital degree of freedom, which are treated for other defect centers in Appendix~\ref{sec:defectmodels}. 
The model discussed here is applicable to an ideal NV center and to implanted group-V donors in silicon, and is referred to as NV-type center. 
In this model, the electron part only contains the quadrupole or zero-field splitting (for electron spin quantum number $s>1/2$) and the Zeeman terms,
\begin{align}
	\label{eq:H_el}
	{H}_{e} &=  {D} {S}_z^2 + \gamma_e {B} {S}_z.
\end{align}

The terms in the nuclear-spin Hamiltonian are analogous to that of ${H}_{e}$, but by contrast to \eref{eq:H_el} are universal to all types of defect centers,
\begin{equation}
	\label{eq:H_nuc}
	H_{n} = Q {I}_z^2 - \gamma_n {B} {I}_z,
\end{equation}
with parameters $Q$ and $\gamma_n$ depending on the nuclear isotope.
Transverse quadrupole components would require symmetry breaking by strain, interfaces, gates etc., and are  neglected here.
Due to the quadrupolar interaction energy $Q$, the nuclear spin levels are not equidistant, which makes them individually addressable~\cite{FernandezdeFuentes2024}.

We consider high-symmetry defects, where the hyperfine tensor is approximately diagonal 
with $A_{xx}=A_{yy}=A_\perp$ and $A_{zz}=A_\parallel$, 
\begin{equation}
	\label{eq:H_hf}
	H_{\rm{hf}} = A_\parallel {S}_z {I}_z + \dfrac{A_\perp}{2} \left( {S}_+ {I}_- + {S}_- {I}_+ \right),
\end{equation}
where $S_{\pm} = S_x \pm i S_y$, $I_{\pm} = I_x \pm i I_y$.
Since the hyperfine coupling between the central spins is mainly due to the Fermi contact interaction, the two components $A_\parallel$ and $A_\perp$ are of the same order of magnitude; for donor spins in silicon, they are often taken as identical~\cite{Morello2020}.

In analogy to the spin-1/2 scheme of Refs.~\cite{Everitt2014,Nemoto2014}, we use the Ising component $A_\parallel$ for the entanglement transfer. The exchange part $A_\perp$ as well as the off-diagonal terms in $H_{e}$ may be suppressed 
by choosing the two electron qubit levels sufficiently separate in energy from neighboring levels in terms of the magnetic quantum number. 
This is usually achieved by tuning $B$ and reduces the undesirable terms to dispersive level shifts. 
We discuss this in more detail for two types of defects in Appendix~\ref{sec:defectmodels}.
In Appendix~\ref{sec:hf_perp} we show explicitly for the NV center in diamond that the influence of this correction on the generated entanglement is very small and may be safely neglected. 
With this secular approximation applied, we are left with an interaction term,
$H_{\rm{hf}} = A_\parallel S_z I_z$.

The nuclear part $H_{n}$ is diagonal and thus commutes with $H_{\rm{hf}}$, which is the only way in which the nuclear spins are operated on. 
This means that $H_{n}$ only generates a single-qudit gate that does not influence the entanglement ${E}$ between remote nuclear spins.
More precisely, if we consider the full hyperfine interaction \eqref{eq:H_hf}, there is a correction which is quadratic in $A_\perp$ but so small that it may be safely neglected.

We assign the electron qubit states $\ket{0}$ and $\ket{1}$ to the magnetic quantum numbers $m_s$ and $m_s'$, respectively, and assume that all other electron levels remain unoccupied throughout.
Thus, we can write the Hamiltonian as $H = \ket{0}\bra{0} \otimes h^{\rm{eff}}_{0} + \ket{1}\bra{1} \otimes h^{\rm{eff}}_{1}$, where $h^{\rm{eff}}_{j}$ are two effective conditional Hamiltonians for the nuclear spin.
If $m_s+m_s'\neq 0$, we can simplify the expressions, without affecting the entanglement, by choosing a symmetrizing nuclear spin rotating frame defined by the time-dependent basis change $U(t)=\exp[-i(m_s+m_s') A_\parallel{I}_z t/2\hbar]$. In this way, we obtain 
\begin{equation}
	\label{eq:Heff}
	h^{\rm{eff}}_{j} = (-1)^{j} (A_\text{net} I_z + \Delta I_0)/2, \quad 
	A_\text{net} \equiv (m_s - m_s') A_\parallel,
\end{equation}
where ${\Delta}$ is the energy difference between levels $m_s$ and $m_s'$.
We assign the qubit states such that $A_\text{net}>0$.
For $s=1/2$ or any two consecutive levels of a higher spin manifold, $A_\text{net} = |A_\parallel|$, while otherwise, 
$A_\text{net}$ is amplified but without any fundamental advantage (i.e., extension of control possibilities) compared to the $s=1/2$ case.
Adding a node index~${K}$, the time-evolution operators corresponding to $h^{\rm{eff}}_{j}$ are 
\begin{align}
	\label{eq:evolop}
	U_{K,j}=\tilde{Z}_{K,(-1)^{j}\varphi_K}, \; \tilde{Z}_{K,\varphi_K}\equiv\exp(-i \varphi_K (I_z+\xi_{K} I_0)/2),
\end{align}
where $\varphi_{K} \equiv A_{\text{net},K}t/\hbar$ and $\xi_{K} \equiv \Delta_{K} / A_{\text{net},K}$.
The dynamics of the electron-nuclear system is thus described by the CPHASE-like gate
$\ket{0}\bra{0} \otimes \tilde{Z}_{K,\varphi_K} + \ket{1}\bra{1} \otimes \tilde{Z}_{K,-\varphi_K}$.
In the hypothetical case $\xi_{K} = 0$, the operator $\tilde{Z}_{K,\varphi_K}$ reduces to the $Z$ rotation (${Z}_{\varphi_K}$).
The electronic level difference ${\Delta}_{K}$ is typically in the GHz regime and may in general cause undesirable rapid oscillations in the entanglement transferred onto the nuclear spins as a function of $\varphi_{K}$. 
Switching to a rotating frame for the electron spin as done in Refs.~\cite{Everitt2014,Nemoto2014} does not remove these oscillations since the measurement needs to be carried out in the laboratory frame. The points of complete deterministic entanglement transfer are unaffected by the oscillations (i.e., independent of the precise value of $\xi_{K}$) due to property (iii) derived in Sec.~\ref{sec:condition_transfer_deterministic}; an additional exception is discussed in the next subsection.

\subsection{Entangling nodes}
\label{sec:el-el-entangling}

For defect centers featuring a spin-photon interface, single-photon interference may be the preferred way of projecting electron spins in distant nodes into an entangled state.
Any other entangling mechanism which is capable of keeping nuclear spin qudits decoupled is explicitly allowed. 
For our purpose it just matters which state $\ket{\phi_{ee}}$ with $E_{ee}=1$ is produced. 
Even though complete deterministic entanglement transfer is unaffected by this choice, it is relevant for other transfer types and for the nuclear spin state being generated. 
Note that while photon-mediated interactions between higher-dimensional spin systems are conceivable~\cite{Tabares2022}, 
in this work we use the electron spins as qubits and correspondingly the spin-photon interface in the traditional binary fashion with two outcomes, e.g., reflection or no reflection of the photon. 

The photonic scheme has been described in Ref.~\cite{Nemoto2014} and analyzed in more detail for both the NV center and a group-IV (silicon) vacancy in diamond in Ref.~\cite{Omlor2025}.
Here we assume a perfect photonic scheme without any losses in the channels and with perfect contrast of the spin-dependent reflection.
We consider the Mach-Zehnder interferometer setup shown in green in \fref{fig:circuit}.
It consists of a single-photon source, a 50:50 beam splitter, the spin-photon interface of each node, another 50:50 beam splitter (which in a two-way setup would coincide with the first one), and a single-photon detector.
We assume that the photon is reflected (with an irrelevant phase shift $\pi$) if the cavity's electron spin (stationary qubit) is in state $\ket{0}$, while it is lost (scattered) for the other qubit state~$\ket{1}$. In neither case a change to the stationary qubit occurs.
Initially, each stationary qubit is in one of the two superposition states $\ket{\pm}=(\ket{0}\pm\ket{1})/\sqrt{2}$.
Without loss of generality, we may consider one of them to be in $\ket{+}$; in particular, we consider the initial state $\ket{+\pm}_{e_a e_b}$.
In the following, we use the notation $\ket{ph_1 ph_2 e_a e_b}$ for the state of both flying and stationary qubits, where `1' and `2' are the photonic channels on the left and right side of the beam splitter, respectively. The (photon amplitude) number $ph_i\in \{0,1\}$ describes the absence/presence of a photon in channel $i$.
A single photon is incindent in channel 1, so the entire initial state is $\ket{10\!+\!\pm}$. 
After scattering from the beam splitter, the state is path-entangled, 
$\frac{1}{\sqrt{2}}(\ket{10\!+\!\pm} +\ket{01\!+\!\pm})$.
The photon modes now interact with the corresponding cavities which yields
$\frac{1}{\sqrt{2}}(\ket{100\pm} +\ket{01\!+\!0})$.
Here we omitted two further terms for which no photon is reflected from either cavity. Effectively this is realized by a post-selection of outcomes with detector click events. 
Now, the reflected photon parts interfere at the second beam splitter, which destroys the which-path information and leads to
$\frac{1}{2}(\ket{100\pm}-\ket{010\pm}+\ket{01\!+\!0}+\ket{10\!+\!0})$. Here, a $\pi$ phase shift occurs for transmission from the upper left to the lower right channel (but could be at any other scattering matrix element).
Finally, by selecting outcomes where the detector at the dark port (channel~2) clicks, the stationary qubits are projected into 
the state $\ket{\Psi^{\mp}}$, where
\begin{align}
	\label{eq:Bell_state_Psi}
	\ket{\Psi^{\pm}} &= \frac{1}{\sqrt{2}}(\ket{01}\pm\ket{10}).
\end{align}
This state can be transformed into any other maximally entangled state by local operations.
In particular, one of the other two Bell states
\begin{align}
	\label{eq:Bell_state_Phi}
	\ket{\Phi^{\pm}} &= \frac{1}{\sqrt{2}}(\ket{00}\pm\ket{11})
\end{align}
is obtained by driving one spin qubit with a $\pi$-pulse. 
On the other hand, driving with a $\pi/2$-pulse yields a two-qubit cluster state, e.g.,
\begin{align}
	\label{eq:cluster_state}
	\ket{\Phi_{\cal C}} &=\frac{1}{\sqrt{2}}(\ket{0+}+\ket{1-}) = 
	\frac{1}{\sqrt{2}}(\ket{\Phi^{-}}+\ket{\Psi^{+}}).
\end{align}
As a side note, if the naturally obtained state $\ket{\Psi^{\pm}}$ is used, the entangled state preparation (excluding initialization) commutes with the electron-nuclear controlled gate, while with driving it does not. 

The probability for the detector to click, i.e., for entangling the remote electron spins, is $1/8$.
Thus the photonic scheme is repeated until success (dash-dotted box in \fref{fig:circuit}) in each iteration of entanglement transfer (dashed box in \fref{fig:circuit}). 
When we give probabilities for the iterative entanglement transfer scheme in Sec.~\ref{sec:schemes}, we do not include the photonic scheme probability, which simply multiplies to the overall probability. 
We further note that an alternative photonic scheme has been proposed~\cite{Omlor2025}, which generates a maximally entangled state deterministically. 
The key difference is that the phase instead of the amplitude of the cavity-reflected photon is dependent on the electron spin,
with the downside of not allowing a quantum non-demolition measurement of the qubit state without making use of an interferometer. 

The photonic scheme described above naturally results in one of the Bell states $\ket{\Psi^{\pm}}$.
Substituting this state along with evolution operators \eqref{eq:evolop} into the effective gate \eqref{eq:Tmat}, one can see that the typically large parameter $\xi$ drops out from the nuclear spin evolution and entanglement, provided $\xi$ is identical for both nodes. For all other states $\ket{\phi_{ee}}$, the generated qudit entanglement typically oscillates rapidly as a function of the CPHASE angle $\varphi$, except at points of complete deterministic transfer (see property (iii) in Sec.~\ref{sec:condition_transfer_deterministic}). 
In Ref.~\cite{Nemoto2014} ($d=2$), the two-qubit cluster state $\ket{\phi_{ee}} = \ket{\Psi_{\cal C}} \equiv (\ket{0+}+\ket{1-})/\sqrt{2}$ was chosen as the resource with the goal of generating a nuclear spin cluster state; here we briefly note that the generated state depends on $\xi$ and is a two-qubit cluster state up to local $Z$ rotations (for any measurement outcome) only for $2\xi$ integer and odd, which could be achieved by fine tuning $B$~\cite{Nemoto2014_comment}.

\subsection{Controlling the hybrid interface}
\label{sec:decoupling}

Since the hyperfine interaction is always on, the operations on the electron spins require a precisely timed schedule. 
In particular, the spin-echo technique is useful for effectively decoupling electron and nuclear spins at all relevant events. 
In principle, this works independently of the nuclear spin quantum number $I$, and hence, we can largely base our timing protocol on Ref.~\cite{Nemoto2014}. 
However, there are some subtle points which are worth to discuss.
We first note that we have not chosen a nuclear spin rotating frame such that $h^{\rm{eff}}_{0}=0$ 
but such that $h^{\rm{eff}}_{0}+h^{\rm{eff}}_{1}=0$, see Sec.~\ref{sec:node_model}. In practice, this does not make any difference, because in either case one needs to precisely record the time electrons spend in a polarized state to track the natural $Z$ evolution of the nuclear spins. 
Denoting the controlled gate $\text{C}(U_{K,0},U_{K,1})$ by $W_K(t)$, we see that
$W_K(t) (Y \otimes I_0) W_K(t) = Y \otimes U_{K,0}U_{K,1}$, where we used that $U_{K,0}$ and $U_{K,1}$ commute.
This shows that the application of a spin-echo $\pi$ pulse on the electron spin at a time~$t$ results in a decoupled evolution of the nuclear spin at the $2t$ time, and further, this holds for arbitrary (even mixed) states of the two electron spins within their qubit subspaces. 
For our particular (symmetrizing) rotating frame $U_{K,0}U_{K,1}=I_0$, so the nuclear spin state at the $2t$~time is identical with that at the beginning of the spin-echo period.

The spin-echo technique is critical during entangling attempts on the electron spins. 
Here it is crucial to synchronize the spin-echo schedule in each node with the operationally relevant events of the entangling mechanism. 
For the case of the photonic entangling scheme described in Sec.~\ref{sec:el-el-entangling}, we suggest defining three time windows as follows, with a spin-echo pulse applied in the center of each window to ensure decoupling of nuclear spins at the end of the window;
this sequence contains only minor modifications with respect to Ref.~\cite{Nemoto2014}.
(i) In each node $K$, the first time window begins with rotating $e_K$ into the superposition state $\ket{\pm}$ and ends at the time $t_{\text{cav},K}$ when the photonic wave part interacts with this cavity. 
(ii) The second window starts at $t_{\text{cav},K}$ and ends at the expected click time $t_\text{click}$ of the detector at the dark port (in the successful case), at which time the electron spins are effectively projected into a new state. Due to the state independence of the spin-echo technique, it is not necessary to make any assumption about the state of the electron spins (e.g., as maximally mixed state in the unsuccessful case~\cite{Nemoto2014}).
(iii) A third window, starting at $t_\text{click}$ and ending at $t_\text{end}$ (defining the end of one internode entangling attempt)  serves as a decision window. 
The decoupling at $t_\text{end}$ is used to proceed with the protocol depending on detector clicks. 
In the successful case (dark port detector clicks), this time commences the natural CPHASE gate between electron and nuclear spins, and the appropriate local gate should be applied at this moment if a state $\ket{\phi_{ee}}$ different from the naturally obtained state $\ket{\Psi^{\pm}}$ is desired. 
The CPHASE gate in node $K$, characterized by the angle $\varphi_K$, is simply a waiting time of length $\hbar\varphi_{K}/A_{\text{net},K}$ with no pulses applied, and extends until the measurement of $e_K$ in the $X$ basis. 
In the unsuccessful case (dark port detector does not click),
one measures and reinitializes the electron spins at the time $t_\text{end}$ to start another entangling attempt. 
Alternatively to the decision window (iii), one could start the CPHASE gate at $t_\text{click}$ in both cases, and choose the CPHASE angle $\varphi$ as the actual one (as a multiple of $2\pi$) in the (un-)successful case.

Errors that accumulate on nuclear spins during attempts to establish
entanglement between remote electron spins have been determined in Ref.~\cite{Nemoto2014} (Sec.~IV of the Supplementary Information therein) as timing issues, nuclear decoherence, and excitation of the electron. 
Here we discuss those error sources in the context of the high-spin memory qudit. 

(i) The first error source is poor timing control and, associated with it, a finite duration of the microwave pulses.
This becomes particularly important for truly high nuclear spins, whose hyperfine coupling is typically an order of magnitude larger than for the NV center. 
While this is beneficial with respect to decoherence, it might make it necessary to drive single-qubit gates by composite pulses or to use optimal control techniques.

(ii) Nuclear spin decoherence. Since during the electron-electron entanglement scheme, nuclear spins are effectively decoupled by  the spin-echo technique, here we consider the `intrinsic' nuclear decoherence due to other noise. This becomes less problematic for the high-spin nuclei, as their coherence times have been shown to be ultralong in experiments. Further, the transfer per iteration is fast as mentioned in (i), and even when the total duration of the scheme, then governed by the time needed to establish electron-electron links, is in the \textmu s range, the error due to intrinsic nuclear decoherence is presumably not dominant.

(iii)
A significant error presumably occurs in the case of a photon being absorbed by the defect. 
In Ref.~\cite{Nemoto2014}, where for the NV center the absorption probability is calculated as smaller than $0.01$, these events are considered as an error source due the possibility of non-spin-conserving transitions back to the ground state, resulting in an error probability of $\sim 3\cdot 10^{-5}$.
However, another error source, which does not seem to be discussed, is the error resulting from the time spent in (unknown) excited states, with typically much stronger hyperfine interaction and without reliable decoupling technique. 
If this error cannot be suppressed, it might be necessary to discard an already established entanglement link in this event, which (together with other photon loss sources) is heralded by the absence of a detector click at both dark and bright ports.

\section{Entanglement accumulation}
\label{sec:schemes}

In this section, we consider two nodes and explore schemes to accumulate entanglement between the nuclear spin qudits via the Ising coupling to their respective electron spins. 
Each node is effectively characterized by the coupling strength $A_\text{net}$ and the transition energy $\Delta$. By default, all nuclear spin levels are used to define the qudit so that $d=2I+1$, but it is possible to deliberately lower the dimension by leaving levels unoccupied at one end (or both ends) of the magnetic ladder.
Then the occupied levels up to a rotating frame evolve equivalently to a nuclear spin with the corresponding level number. 
This gives the opportunity to define from nuclear spin $I$ a qudit with dimension $d=2^{\lfloor \log_2 {(2I+1)} \rfloor}$, which, as shown in Sec.~\ref{sec:scheme_deterministic}, is suitable for deterministic entanglement accumulation. As an example, a ${d}=8$ qudit may be defined from the $I=9/2$ spin of \isotope{Ge}{73} by leaving the two levels $m_{I}=\pm 9/2$ empty. The only disadvantage is possible leakage into the unoccupied levels especially during driving operations, causing lower fidelities. 
An alternative but more profound modification is defining a qudit from the electron spin (provided its quantum number is $s>1/2$), which, for the same coupling $S_z I_z$, would alter the set of favorable qudit dimensions ${d}$ accordingly, since the base two in $2^{n}$ is the qubit dimension. 

Since the rotation axis is fixed throughout and no other (driving) operations are applied, the effective gates ($\{{\cal T}^{(\nu)}\}$) commute with each other, so that generally the order in which the phases $\{\varphi^{(\nu)}\}$ are applied does not affect the final state.
The rank doubles in each iteration ($r^{(\nu)}=2r^{(\nu-1)}$) or reaches the maximum ($r^{(\nu)}=d$), whenever we apply a phase $\varphi^{(\nu)}$ that has not been applied in one of the previous iterations; otherwise the rank remains the same or even decreases.
(If $\ket{\phi_{ee}^{(\nu)}}\neq \ket{\phi_{ee}^{(\nu')}}$, then only discrete values of $\varphi^{(\nu)}=\varphi^{(\nu')}$ do not double the rank.)

The state $\ket{\psi^{(\nu)}_{J}}$ after $\nu$ iterations depends on the $2\nu$ measurement outcomes ${J} \equiv \{j_a^{(\mu)},j_b^{(\mu)}|\mu=1,\dots,\nu\}$ and its entanglement $E^{(\nu)}_{J}$ may in general also depend on the measurement record $J$. 
Of interest is the expectation value of the entanglement,
$\langle E^{(\nu)} \rangle = \sum_{J} P_{J} E^{(\nu)}_{J}$,
with $P_{J}=\braket{\psi^{(\nu)}_{J}|\psi^{(\nu)}_{J}}$.
Without giving a proof, we note that $\langle E^{(\nu)} \rangle$ is $2\pi$-periodic in any $\varphi^{(\nu)}$ and for any $\xi$.
In the following, we omit the measurement outcome index $J$ for ease of notation.

\subsection{Constructed deterministic entanglement transfer scheme}
\label{sec:scheme_deterministic}

\begin{figure}[t]
	\centering
	\includegraphics[width=\figwidth]{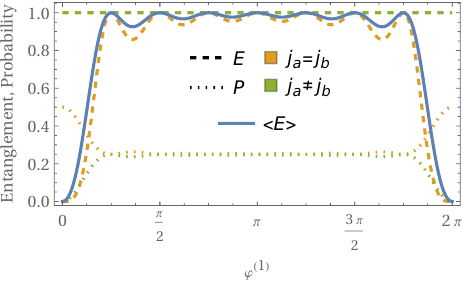}
	\caption{First iteration of entanglement transfer (${d}=8$, $\ket{\phi_{ee}}=\ket{\Psi^{+}}$), showing entanglement and probability for individual measurement outcomes $(j_a,j_b)$ as well as the expected entanglement, as functions of the CPHASE angle. 
	}
	\label{fig:E_round1}
\end{figure}

A unique property of quantum spin is that the rotation around a fixed axis can generate a complete orthonormal set of states from a single, suitably chosen, state vector. 
More specifically, for the essential $z$-axis rotation $\tilde{Z}_{K,\varphi_{K}}$, \eref{eq:evolop},
the necessary and sufficient requirement is that the initial state is uniform in the eigenbasis of $I_z$
(i.e., each component has square modulus $1/d$), and the rotation angles form the cyclic group, $\varphi_k = 2\pi k / d$, with integer $k$. 
We can thus exploit a conditional $I_z$ rotation to implement a simple realization of repeated complete deterministic entanglement transfer. 
For this, the initial nuclear two-qudit state needs to be chosen as 
\begin{equation}
	\label{eq:psi0}
	\ket{\psi^{(0)}} = \ket{+_{d}}\otimes\ket{+_{d}}, \quad
	\ket{+_{d}}\equiv \frac{1}{\sqrt{d}} \sum_{i=0}^{{d}-1}\ket{i}.
\end{equation}

For the first iteration, the orthogonality condition \eqref{eq:cond_complete} is
$\bra{+_{d}} \tilde{Z}_{K,2\varphi^{(1)}_K} \ket{+_{d}} = 0$
and yields $\varphi^{(1)}_K={k} 2\pi/{d}$ with integer ${k}\in \{1,\dots,{d}-1\}$. 
Choosing $\varphi_a^{(\nu)}=\varphi_b^{(\nu)} \equiv \varphi^{(\nu)}$, the plot in 
\fref{fig:E_round1} confirms for ${d}=8$ and $\ket{\phi_{ee}}=\ket{\Psi^{+}}$ that the expected entanglement reaches 1~ebit for these angles.
The individual probabilities $P_{j_a j_b}$ and entanglements $E_{j_a j_b}$ are also shown; note that $E_{0 1}=E_{1 0}=1$ for any $\varphi^{(1)}$.
The plots for other ${d}$ are analogous.

For any of the later iterations $\nu$ we use that the transfer of 1~ebit has been deterministically achieved in each previous iteration by choosing suitable angles $\{\varphi^{(\mu)}\}$, so that the Schmidt rank of $\ket{\psi^{(\nu-1)}}$ is $r^{(\nu-1)}=2^{\nu-1}$. 
This further implies that for each node ${K}$, the $2^{\nu-1}$ states,
\begin{equation}
	\ket{{v}_{\{\sigma^{(\dots)}\}}^{(\nu-1)}}_{K} = \tilde{Z}_{\sigma^{(\nu-1)}\varphi^{(\nu-1)}} \dots \tilde{Z}_{\sigma^{(1)}\varphi^{(1)}} \ket{+_{d}},
\end{equation}
with sign factors $\sigma^{(\mu)}=\pm 1$ are an orthonormal set
with the same span as the Schmidt basis for $\ket{\psi^{(\nu-1)}}$. 
Thus we can express the single-node condition \eqref{eq:cond_complete} as
\begin{equation}
	\label{eq:cond_spinrot_complete}
	\!\!{\phantom{|}}_{K}\!\bra{{v}_{\{\sigma^{(\dots)}\}}^{(\nu-1)}} \tilde{Z}_{K,2\varphi^{(\nu)}} \ket{{v}_{\{\sigma^{(\dots)}\}}^{(\nu-1)}}_K = 0,
\end{equation}
which must hold for any of the $2^{2\nu-2}$ combinations of sign factors.
Since the parameter $\xi_K$ only causes scalar phase factors, it drops out from the condition as expected for a complete deterministic transfer.
Explicit evaluation yields
\begin{align}
	\label{eq:criterion}
	\sum_{k=-I}^{I} \exp\left[-ik\left(\varphi^{(\nu)}+\sum_{\mu=1}^{\nu-1}\tau^{(\mu)}\varphi^{(\mu)}\right)\right] = 0,
\end{align}
for all combinations of $\{\tau^{(\mu)}\}$, where  $\tau^{(\mu)} \in \{-1,0,1\}$.
With $\tau^{(\mu)}=0$ for all $\mu$, we obtain the same phase discretization as for the first iteration, $\varphi^{(\nu)}=k^{(\nu)} 2\pi/{d}$ with integer $k^{(\nu)}\in \{1,\dots,{d}-1\}$. 
The other combinations further restrict the allowed indices $k^{(\nu)}$.
For the second iteration, we find the constraints $k^{(2)}\neq k^{(1)}$ and $k^{(2)}\neq d-k^{(1)}$.
Any number up to $\nu_{\text{max}} = \lfloor E_{d} \rfloor$ of maximal transfer iterations can be realized by the set of indices $k^{(\nu)} = 2^{\nu-1}$, which is also the set with the shortest gate duration.
Thus, with each $\ket{\phi_{ee}^{(\nu)}}$ providing 1~ebit, the phase set
\begin{equation}
	\label{eq:phaseset_dt}
	\varphi^{(\nu)}=\frac{2^{\nu}\pi}{d},\quad \nu=1,\dots,\nu_{\text{max}}, \quad \nu_{\text{max}} = \lfloor E_{d} \rfloor
\end{equation}
deterministically results in $E = \lfloor E_{d} \rfloor$, which corresponds to the maximal entanglement $E_{d}$ for the case $d=2^n$ with any integer $n$.
The order in which the phases $\{\varphi^{(\nu)}\}$ are applied, does not matter, and could also differ between the two nodes, but this would result in a longer total duration. 

\begin{figure}[t]
	\centering
	\includegraphics[width=\figwidth]{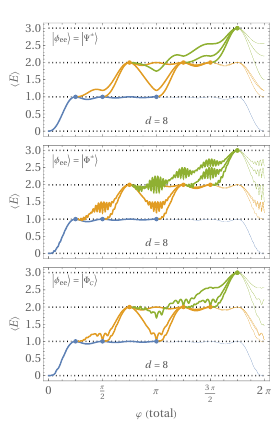}
	\caption{Generating maximal entanglement deterministically in three iterations for ${d}=8$. The expected entanglement without postselection is shown as a function of the accumulated CPHASE angle. Only the fastest phase combinations are shown (set~\eqref{eq:phaseset_dt}, any ordering). 
	}
	\label{fig:E8d}
\end{figure}

The plots in \fref{fig:E8d} confirm that this phase set indeed leads to a deterministic increase of the entanglement by 1~ebit in each iteration, for all choices of (maximally entangled) $\ket{\phi_{ee}}$. 
This deterministic increase is possible until $E = \lfloor E_{d} \rfloor$ is reached. 
The total CPHASE angle required to achieve this is smaller than $2\pi$ (e.g., $\varphi=7\pi/4$ for ${d}=8$ shown in \fref{fig:E8d}).

Not surprisingly, the bipartite entanglement generated by such a series of 1-ebit transfers is decomposable~\cite{Kraft2018}, i.e., the correlations are reproducible by $n$ pairs of qubit Bell pairs. 
This is not the case for the remaining schemes based on incomplete transfers, nor does it generalize to the multipartite case (Sec.~\ref{sec:multipartite}).
Furthermore, this property should not necessarily be considered a downside for quantum computing tasks.

For qudits with $d\neq 2^n$ we require at least ${\nu_{\text{max}}=\lceil E_{d} \rceil}$ iterations to reach $E=E_{d}$, but the constraints implied by \eref{eq:criterion} cannot be fulfilled for any further iteration, ${\nu>\lfloor E_{d} \rfloor}$. 
It further turns out that, in agreement with the deviating constraints for generating maximal entanglement deterministically (Sec.~\ref{sec:condition_Emax_deterministic}), after such $\lfloor E_{d} \rfloor$ iterations of deterministic transfer we are in a bad position for further increasing the expected entanglement significantly (e.g., for some $d$ the entanglement directly decreases from this point). 
Reaching higher values of entanglement for such qudit dimensions requires other schemes, which are identified in the following subsections; in particular, we don't find a method without driving that achieves (near-) maximal entanglement deterministically.

\subsection{Constructed entanglement transfer scheme ($d=3$)}
\label{sec:scheme_deterministic_3d}

We now consider the lowest unfavorable qudit dimension, $d=3$, and investigate whether maximal entanglement can be generated deterministically in some other way. 
A scheme for $d=3$ might also be useful as an elementary block of a scheme for (even) multiples of $d=3$.
It is plausible to follow a reverse engineering approach by first applying the constraints derived in Sec.~\ref{sec:condition_Emax_deterministic} for the final iteration.
For $d=3$, the controlled gate $\text{C}(U_{K,0},U_{K,1})$ is described by
\begin{equation}
	U_{K,0}\hc U_{K,1} = e^{i\xi_K \varphi_K} \begin{pmatrix}
		e^{i\varphi_K} & 0 & 0 \\ 0 & 1 & 0 \\ 0 & 0 & e^{-i\varphi_K}
	\end{pmatrix}.
\end{equation}

We denote the index of the final iteration by $\nu_\text{f}$.
Due to the constraints on the Schmidt coefficients (item~2 from Sec.~\ref{sec:condition_Emax_deterministic}), we have
\begin{align}
	S_\downarrow^{(\nu_\text{f}-1)} &= \begin{pmatrix}
		\frac{1}{3}+\varepsilon & 0 & 0 \\ 0 & \frac{1}{3}& 0 \\ 0 & 0 & \frac{1}{3}-\varepsilon
	\end{pmatrix}
\end{align}
with $0<\varepsilon\le 1/3$. Only single element blocks are present, and thus, unitary ${\cal U}_K$ has the antidiagonal structure,
\begin{equation}
	{\cal U}_K = \begin{pmatrix}
		0 & 0 & e^{i\alpha_+} \\ 0 & e^{i\alpha_0} & 0 \\ e^{i\alpha_-} & 0 & 0
	\end{pmatrix},
\end{equation}
with yet arbitrary phases.
Since $U_{K,0}\hc U_{K,1}$ and ${\cal U}_K$ are related by the unitary transformation $Q_K$, their eigenvalues must be identical. 
The three eigenvalues of ${\cal U}_K$ are $\{\pm e^{i(\alpha_++\alpha_-)/2}, e^{i\alpha_0}\}$.
There are two solutions, namely
\begin{align}
	\label{eq:UK_sol1}
	{\cal U}_K &= e^{i\xi_K \pi/2} \begin{pmatrix}
		0 & 0 & e^{i\alpha_K} \\ 0 & 1 & 0 \\ -e^{-i\alpha_K} & 0 & 0
	\end{pmatrix}\land\ \varphi_K = \frac{\pi}{2},
\end{align}
and
\begin{align}	
	\label{eq:UK_sol2}
	{\cal U}_K &= e^{i\xi_K \pi} \begin{pmatrix}
		0 & 0 & e^{i\alpha_K} \\ 0 & -1 & 0 \\ e^{-i\alpha_K} & 0 & 0
	\end{pmatrix}\land\ \varphi_K = \pi ,
\end{align}
each containing one arbitrary phase angle $\alpha_K$.
So far we can choose any combination of the two solutions for the two nodes.
The remaining constraints from Sec.~\ref{sec:condition_Emax_deterministic} are $E_{ee}^{(\nu)}=1$
and Eqs.~\eqref{eq:cond_gen3}.
The constraints from Eqs.~\eqref{eq:cond_gen3} are least restrictive if we choose $\varepsilon=1/3$. As we will see, this still fixes the state from the previous iteration (up to a phase factor), so that it is not necessary to evaluate the case $\varepsilon<1/3$, in which at least one additional iteration would be required ($\nu_\text{f}\ge 3$) because $r^{(\nu_\text{f}-1)}=3$, which only lowers the success probability.
Thus, we now take $\varepsilon=1/3$.
The upper left and lower right element (block) on the l.h.s.~of both Eqs.~\eqref{eq:cond_gen3} then vanish, which yields $P^{(\nu_\text{f})}_\text{eq}=1/2$.
The central element of the two solutions for ${\cal U}_K$ is $\pm e^{i\xi_K \varphi_K}$.
Since this is a scalar, the l.h.s.~of \eref{eq:cond_gen3_a} also vanishes for the central block. The remaining condition from the central block of \eref{eq:cond_gen3_b} becomes
\begin{equation}
	\label{eq:cond_c}
	c^2 \cos{(\xi_a \varphi_a + \xi_b \varphi_b)}
	-
	(\tfrac{1}{2}-c^2) \cos{(\xi_a \varphi_a - \xi_b \varphi_b)} = 0.
\end{equation}
Depending on the values of $\xi_a$ and $\xi_b$, this equation typically has zero or one solution for $c$ within the allowed range ($0\le {c} \le 1/\sqrt{2}$), except at discrete points where all allowed $c$ yield a solution. 
Below we will give an argument for choosing the solution \eqref{eq:UK_sol1} in both nodes, so $\varphi_a=\varphi_b=\pi/2$; in this case, the discrete points are at: one $\xi_K$ integer and even, and the other $\xi_K$ integer and odd.
For simplicity we now take $\xi_a = \xi_b \equiv \xi$, which seems to be the most relevant practical case. 
In this case (and still $\varphi_a=\varphi_b=\pi/2$), \eref{eq:cond_c} has one (and only one) solution if $\xi$ is not farther away from an even number than $1/2$ (i.e., $|\xi\,\text{mod}\,2-1|\ge 1/2$). If we consider integer and even $\xi$, then the solution is $c=1/2$, which means that $\ket{\phi_{ee}^{(2)}}$ is a two-qubit cluster state. 

The chosen solutions (out of \eqref{eq:UK_sol1} and \eqref{eq:UK_sol2}) for the two nodes are sufficient to essentially fix the state $\ket{\psi^{(\nu_\text{f}-1)}}$ prior to the final iteration.
From \eref{eq:Schmidt_QK}, $\psi^{(\nu_\text{f}-1)} = Q_a \sqrt{S_\downarrow} Q_b\tran$.
Since ${U}_{K} = {Q}_{K} {\cal U}_{K} {Q}_{K}\hc$ is diagonal, the columns of ${Q}_{K}\hc$ are the ordered eigenvectors of ${\cal U}_{K}$.
The effective gate ${\cal T}_{j_a j_b}$ is also diagonal, 
which allows us to inspect the possibility of reaching the state $\ket{\psi^{(1)}}$ from a product state $\ket{\psi^{(0)}}$ in the first iteration. 
(It is fairly obvious that using more than one iteration to generate $\ket{\psi^{(\nu_\text{f}-1)}}$ must further lower the success probability, hence we now set $\nu_\text{f} = 2$.)
It turns out that the structure of $\ket{\psi^{(1)}}$ (in particular, the positions of the vanishing elements) is fitting to the model only if we choose the solution \eqref{eq:UK_sol1} in both nodes, and this choice yields
\begin{equation}
	\label{eq:psi1_recon}
	\psi^{(1)} = \frac{1}{\sqrt{6}}
	\begin{pmatrix}
		-1 & 0 & 1 \\ 0 & \sqrt{2}e^{i(\alpha_a+\alpha_b)} & 0 \\ 1 & 0 & -1
	\end{pmatrix}.
\end{equation}
Despite the phase freedom at the central element, the state~$\ket{\psi^{(1)}}$ seems too strongly constrained to be capable of being prepared by entanglement transfer.
However, it turns out that we can prepare this state by a single entanglement transfer followed by postselection with a success rate of $0.5$ (next paragraph).
Alternatively, we can aim to prepare its Schmidt coefficients by entanglement transfer and apply local basis changes afterwards; this is shown to allow for deterministic schemes (paragraph below).

Instead of performing a general construction, we use the observation that Bell states are the suitable resource to directly generate the state given in \eref{eq:psi1_recon} in the first iteration.
More specifically, we choose $\ket{\phi_{ee}^{(1)}}=\ket{\Psi^+}$, select ${\varphi_a^{(1)}=\varphi_b^{(1)}=\pi}$, and postselect $j_a=j_b$; other options are given below.
It is convenient to partition the $d^2$ diagonal elements of ${\cal T}_{j j}$ into a $d$ by $d$ matrix, explicitly
\begin{equation}
	\label{eq:Tjjmat}
	[\diag({\cal T}_{j j})]_\text{mat} = \frac{1}{\sqrt{2}}
	\begin{pmatrix}
		1 & 0 & -1 \\ 0 & 1 & 0 \\ -1 & 0 & 1
	\end{pmatrix},
\end{equation}
so that $\psi^{(1)} \sqrt{P_{jj}} =[\diag({\cal T}_{j j})]_\text{mat} \odot \psi^{(0)}$, where $\odot$ is elementwise multiplication, and $\psi^{(\nu)}$ are normalized.
The effective gate described by \eref{eq:Tjjmat} has a compatible structure with the state from \eref{eq:psi1_recon}. In particular, 
the state $\ket{\psi^{(1)}}/2$ (with norm $1/2$) is generated by  ${\cal T}_{j j}$ from the product state
\begin{equation}
	\label{eq:scheme_3d_init}
    \psi^{(0)} = \frac{1}{2\sqrt{3}}\begin{pmatrix}
		1 & -\sqrt{2}e^{i(\alpha_a+\alpha_b)} & 1 
	\end{pmatrix}
	\otimes
	\begin{pmatrix}
		1 \\ 1 \\ 1
	\end{pmatrix}.
\end{equation}
Since both outcomes with $j_a=j_b$ are allowed, resulting in the same state up to an irrelevant global phase, the success probability is $1/2$. 
We can summarize the entire scheme, with the parameter~$\xi$ tuned to an even integer number, as follows,
\begin{subequations}
	\label{eq:scheme_3d_prob}
    \begin{align}
    	&\ket{\phi_{ee}^{(1)}} = \ket{\Psi^{+}}, \quad \varphi_a^{(1)}=\varphi_b^{(1)}=\pi, \quad j_a^{(1)}=j_b^{(1)}, \\
    	&\ket{\phi_{ee}^{(2)}} = \ket{\Phi_{\cal C}} , \quad \varphi_a^{(2)}=\varphi_b^{(2)}=\frac{\pi}{2}.
    \end{align}
\end{subequations}
This is the driving-free scheme with the highest overall success rate (equal to 1/2) for generating $d=3$ maximal entanglement. 
We could equivalently choose
$\ket{\phi_{ee}^{(1)}} = \ket{\Psi^{-}}$ and postselect $j_a^{(1)}\neq j_b^{(1)}$; further, with $\xi$ tuned to an integer, $\ket{\phi_{ee}^{(1)}} = \ket{\Phi^{\pm}}$ would also work.

The second pursuable option is generating the required Schmidt coefficients $\{\sqrt{2/3},\sqrt{1/3}\}$ in the first iteration with arbitrary Schmidt vectors, and use local driving of nuclear spins to apply the required state correction afterwards. 
One way to achieve such Schmidt coefficients deterministically can be deduced from our transferability condition for complete transfer. From the previous subsection (Sec.~\ref{sec:scheme_deterministic}) we know that with nuclear spins initialized in the equal superposition state \eqref{eq:psi0}, a complete deterministic transfer takes place at $\varphi_a^{(1)}=\varphi_b^{(1)}=2\pi/3$, irrespective of the two-qubit resource state $\ket{\phi_{ee}^{(1)}}$. If we choose a partially entangled state $\ket{\phi_{ee}^{(1)}}$ with just the required Schmidt coefficients $\{\sqrt{2/3},\sqrt{1/3}\}$, and apply the transfer with these settings, the obtained state $\ket{\psi^{(1)}_{j_a j_b}}$ must have those Schmidt coefficients for any measurement outcome. The advantage of this solution is that it works independently of the value of~$\xi$; however, one has to find a way for preparing a partially entangled state $\ket{\phi_{ee}^{(1)}}$.
In the case of a maximally entangled state $\ket{\phi_{ee}^{(1)}}$, we have been able to determine a solution by numerical evaluation only. 
Without going into details, this solution consists of choosing the default initial state \eqref{eq:psi0}, $\ket{\phi_{ee}^{(1)}} = \ket{\Phi_{\cal C}}$, one out of suitable values for $\varphi_a^{(1)}=\varphi_b^{(1)}$ (e.g., $\approx$1.71255), and fine-tuning the parameter~$\xi$ (to an integer multiple of $\approx$1.83543).

So far in this subsection, we have identified how to realize the deterministic transfer represented by the truncated triangle in \fref{fig:transferability}~(b), and further a deterministic (complete or incomplete) transfer reaching the base level $E=-\frac{2}{3}+\log_2 3$ of this triangle in one iteration from $E=0$. 
The remaining question is whether these schemes are useful to construct schemes for qudits with dimension $d$ given by a multiple of three, in particular $d=6$, where one complete transfer, indicated by the triangle starting with rank $3$ in \fref{fig:transferability}~(a), from $E=\log_2 3$ would lead to maximal entanglement. 
A solution to this question is obtained by virtually reducing the dimension from six to three in the first iterations by leaving three outermost levels unoccupied during initialization. 
This allows us to generate in $d=6$ the Schmidt coefficients $\{\sqrt{1/3},\sqrt{1/3},\sqrt{1/3}\}$ via any one of the schemes from above; 
in particular, a deterministic scheme can be applied, even though driving operations have to be translated to account for the higher-dimensional space. 
Identifying a complete transfer from this point (i.e., from ${r^{(2)}=3}$ to ${r^{(3)}=6}$) is relatively simple based on the condition in \eref{eq:cond_complete}. As an example, we can set $\varphi=\pi$ (i.e., $\varphi_a^{(3)}=\varphi_b^{(3)}=\pi$), which gives $U_{K,0}\hc U_{K,1} \sim \diag(\{1,-1,1,-1,1,-1\})$. Thus, the Schmidt vectors in both nodes could be chosen as the columns of the matrix
$V_K = \mathds{1}_3 \otimes (1, 1)\tran$,
independently of $\xi$ and the choice of EPR pair $\ket{\phi_{ee}^{(3)}}$.
Since any other set of three Schmidt vectors (in the higher-dimensional space, $d=6$) are related to this set by a unitary transformation, there is a driving operation that facilitates the required basis change of the nuclear state after the second iteration.
Therefore, a deterministic scheme (or a driving-free scheme with success rate 0.5) is also available for $d=6$.

To summarize, in this subsection, we have identified schemes for maximally entangling qutrit pairs via two successive transfers from the electron spin qubits. 
A scheme without driving \eqref{eq:scheme_3d_prob} based on an unconventional initial state \eqref{eq:scheme_3d_init} (unequal superposition in one node) is available with a success rate of 50\% due to the first iteration. 
The inclusion of nuclear spin driving in between both iterations allows for deterministic schemes. 
All schemes can be adopted to $d=6$ by combining with a complete 1-ebit transfer without reducing the success rate. 
Even though other multiples such as $d=9$, or prime numbers other than $d=3$, are expected to require individual strategies, we may infer from the obtained results that intermittent nuclear spin driving allows the construction of deterministic schemes for generating maximal entanglement between qudits of any dimension $d$. 

\subsection{Other probabilistic schemes and optimization}
\label{sec:scheme_prob}

In this subsection, we return to a simpler setting and explore some further schemes for probabilistic entanglement accumulation.
In particular, we exclude nuclear-spin driving, consider for the initial state $\ket{\psi^{(0)}}$ the equal superposition state from \eref{eq:psi0}, and further choose the same resource state $\ket{\phi_{ee}}$ in all iterations.
The arguments for these choices are a possibly easier implementation, as well as a simple generalizability to the multipartite case as discussed in Sec.~\ref{sec:multipartite}.
For $d\neq 2^n$ these restrictions result in less optimal bipartite schemes than the scheme without driving constructed in the previous subsection. 
Apart from that, for simplicity and because no benefit has been obtained so far from a different choice, we still consider identical nodes ($\xi_a = \xi_b \equiv \xi$) and apply the same CPHASE angle to both nodes, $\varphi_a^{(\nu)}=\varphi_b^{(\nu)}\equiv\varphi^{(\nu)}$.

While it is possible to construct the schemes in this setting from general conditions,  
for simplicity we identify them by inspection of the output states together with the numerical evaluation of the entanglement. 
Due to the projecting character of the effective gate ${\cal T}^{(\nu)}_{j_a j_b}$, it is of no advantage to use more than the minimum number of iterations required for generating $E=E_{d}$; thus, in this subsection $\nu_{\text{max}}=\lceil E_{d} \rceil$.

From \eref{eq:Tmat} and \eref{eq:evolop} we know that ${\cal T}^{(\nu)}_{j_a j_b}$ is diagonal, and (in the case $\xi=0$) the diagonal elements are linear combinations of at most four terms of the form
$\exp(i {m} \varphi^{(\nu)} /2)$ with integer ${m}$ and $|m|<d$.
For certain discrete phase sets, the output states take a simpler form than for arbitrary angles; e.g., the scheme \eqref{eq:phaseset_dt} consists of power-of-two multiples of~$\pi/d$. 
Another universal phase set consists of power-of-two fractions $2\pi/2^n$, differing from the previous case for $d\neq 2^n$. For these angles, the number of different values of $E$ over the set of $4^{\nu_{\text{max}}}$ possible measurement outcome series becomes particularly small. 

To see that this phase set is also suitable for the probabilistic generation of maximal entanglement, let us start by considering Bell states as the entanglement source, for simplicity $\ket{\phi_{ee}}=\ket{\Psi^{+}}$, and postselecting $j_{a}^{(\nu)}=j_{b}^{(\nu)}$ in each iteration~$\nu$. For this choice, the state $\ket{{{\psi}}^{(\nu)}}$ has the following coefficients in the computational basis,
\begin{equation}
	\left({{\psi}}^{(\nu)}\right)_{i_a,i_b} = \braket{i_a,i_b|{{\psi}}^{(\nu)}} = \frac{1}{2^{\nu/2}{d}} \prod_{\mu=1}^{\nu}\cos\left(\frac{i_a-i_b}{2}\varphi^{(\mu)}\right).
\end{equation}
Now, we choose the phase set $\{\varphi^{(\nu)}\}$ ($\nu=1,\dots,\nu_{\text{max}}$) with
\begin{equation}
	\label{eq:phaseset_pt}
	\varphi^{(\nu)}=\frac{2\pi}{2^{\nu}}(2p^{(\nu)}+1), \quad \nu_{\text{max}} = \lceil E_{d} \rceil,
\end{equation}
where $p^{(\nu)}$ are arbitrary non-negative integers and are typically chosen as $p^{(\nu)}=0$.
With this choice, we find that the state becomes a two-qudit Bell state,
${{\psi}}^{(\nu)}= \mathds{1}_d/d$ with norm~$1/\sqrt{d}$.
For $d\neq 2^n$, any other measurement outcome series (i.e., measuring $j_{a}^{(\nu)}\neq j_{b}^{(\nu)}$ in one or more iterations) results in a partially entangled state. Thus, the scheme~\eqref{eq:phaseset_pt} generates $E=E_{d}$ with success rate $P=1/{d}$ for such ${d}$, while coinciding with the deterministic scheme~\eqref{eq:phaseset_dt} for $d= 2^n$. 

\begin{figure}[t]
	\centering
	\includegraphics[width=\columnwidth]{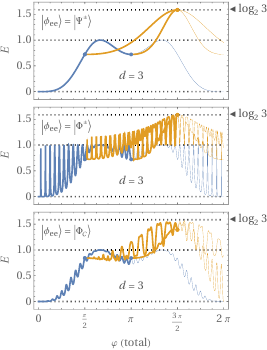}
	\caption{Generating maximal entanglement in two iterations for ${{d}=3}$, using a postselection of $j_{a}^{(\nu)}=j_{b}^{(\nu)}$ in each iteration~$\nu$. 
	Different panels correspond to different entangled electron states as indicated. 
	The electron qubit splitting parameter is chosen to be $\xi=20$.
	}
	\label{fig:Eps3d}
\end{figure}

The top panel of \fref{fig:Eps3d} shows the accumulation of entanglement for ${d}=3$ based on the phase set \eqref{eq:phaseset_pt} for the state $\ket{\phi_{ee}}=\ket{\Psi^{\pm}}$, independently of $\xi$. 
The non-deterministic entanglement transfer is generally dependent on the choice of~$\ket{\phi_{ee}}$, and it is worth exploring other states. 
For states other than  $\ket{\Psi^{\pm}}$, oscillating modulations appear, typically with period $\pi/\xi$.
For $\ket{\phi_{ee}}=\ket{\Phi^{\pm}}$, exact maximal entanglement is reached only for $\xi$ being an integer multiple of $2^{\nu_{\text{max}}-1}$ (i.e., $\xi$ integer and even for ${d}=3$), and the success rate is the same as for $\ket{\phi_{ee}}=\ket{\Psi^{\pm}}$. 
For $\ket{\phi_{ee}} = \ket{\Phi_{\cal C}}$, on the other hand, only near-maximal entanglement is reached, but the success rate can be higher as discussed in the following. 

We observe that for any $d\ge 2$, the protocol \eqref{eq:phaseset_pt} is the only one generating maximal entanglement exactly and with finite probability, and for $d\neq 2^n$ this requires $\ket{\phi_{ee}}$ to be a Bell state. 
However, if we relax our goal from maximal to near-maximal entanglement, there might be other schemes with a possibly higher success rate. 
As an example, consider the case $\ket{\phi_{ee}} = \ket{\Phi_{\cal C}}$ and $d=3$ with the postselection from above (lowest panel of \fref{fig:Eps3d}), for which we numerically find the optimal phase set $(\pi, 0.575\pi)$ with $E/E_{d}\approx 0.989$ and a success probability $P\approx 0.258$. It turns out that in this particular case we can drop the postselection in the first iteration and still obtain exactly the same entanglement $E=E(\varphi^{(1)}=\pi,\varphi^{(2)})$; the phase set above (where the order now matters) then leads to the same $E/E_{d}\approx 0.989$ with a success probability twice as large, $P\approx 0.516$.

A similar numerical optimization is also possible for higher $d\neq 2^n$, but depends on where the weight is put in the tradeoff between entanglement~$E$ and success rate~$P$.  
From an alternative viewpoint, it could also be relevant to optimize the expected entanglement $\langle E \rangle$ averaged over all measurement outcomes.
Interestingly, we find that for this purpose, $\ket{\phi_{ee}} = \ket{\Phi_{\cal C}}$ is the optimal source state (for a properly tuned parameter $\xi$), and the optimal phase set is essentially just the set~\eqref{eq:phaseset_pt} (with at most marginal deviations). 
Before we discuss this sort of optimization in more detail, it is thus worth inspecting the set of entangled states generated by the phase set~\eqref{eq:phaseset_pt}.
A statistical evaluation is shown for up to $d=16$ in Table~\ref{tab:scheme1}. 
As for the postselected results discussed above, the expected entanglement for this phase set is periodic in $\xi$ with maxima at integer multiples of $2^{\nu_{\text{max}}-1}$; note further that for phase set~\eqref{eq:phaseset_pt} and such values of $\xi$, the Bell states $\ket{\Phi^{\pm}}$ and~$\ket{\Psi^{\pm}}$ are equivalent in terms of the generated entanglement.
The grey rows indicate the dimensions ${d}=2^n$, for which the scheme coincides with \eqref{eq:phaseset_dt} and hence works deterministically. 
The worst ratio $\langle{E}\rangle/E_{d}$ is typically found for the dimensions directly above each $2^n$, with an increase towards the next perfect row, ${d}=2^{n+1}$.
Comparing Bell and two-qubit cluster states, we find that the latter allows higher~$\langle{E}\rangle$, with a weaker dependence of the entanglement on the measurement outcomes. The cluster state is also the best for optimizing $\langle E \rangle$ by adapting the CPHASE angles as discussed in the remainder. 

\begin{table}[b]
	\centering
	\begin{tblr}{
		colspec={|ccc|cccc|cccc|}	,
		colsep=2pt,
		row{3}={bg=gray!30},
		row{5}={bg=gray!30},
		row{9}={bg=gray!30},
		row{17}={bg=gray!30},
		rowsep=0pt,
		row{1}={rowsep=3pt},
		row{2}={rowsep=2pt},
	}
		\toprule
		& & & \SetCell[c=4]{l}{$\ket{\Phi^{\pm}}$ or $\ket{\Psi^{\pm}}$} & & & & \SetCell[c=4]{l}{$\ket{\Phi_{\cal C}}$}  & & & \\	
		\midrule
		${d}$ & $\nu_{\text{max}}$ & ${E}_{d}$ & $\langle{E}\rangle$ & $\frac{\langle{E}\rangle}{{E}_{d}\%}$ & \#$(E)$ & $\sigma({E})$ & $\langle{E}\rangle$ & $\frac{\langle{E}\rangle}{{E}_{d}\%}$ & \#$(E)$ & $\sigma({E})$ \\
		\midrule
		2 & 1 & 1 & 1 & 100 & 1 & 0 & 1 & 100 & 1 & 0 \\
		3 & 2 & 1.585 & 1.195 & 75.4 & 2 & 0.276 & 1.392 & 87.8 & 1 & 0 \\
		4 & 2 & 2 & 2 & 100 & 1 & 0 & 2 & 100 & 1 & 0 \\
		5 & 3 & 2.322 & 1.881 & 81.0 & 5 & 0.354 & 2.065 & 88.9 & 3 & 0.077 \\
		6 & 3 & 2.585 & 2.274 & 88.0 & 4 & 0.237 & 2.372 & 91.8 & 4 & 0.106 \\
		7 & 3 & 2.807 & 2.617 & 93.2 & 2 & 0.078 & 2.676 & 95.3 & 1 & 0 \\
		8 & 3 & 3 & 3 & 100 & 1 & 0 & 3 & 100 & 1 & 0 \\
		9 & 4 & 3.170 & 2.765 & 87.2 & 9 & 0.432 & 2.934 & 92.6 & 10 & 0.075 \\
		10 & 4 & 3.322 & 2.906 & 87.5 & 9 & 0.329 & 3.055 & 92.0 & 12 & 0.079 \\
		11 & 4 & 3.459 & 3.072 & 88.8 & 9 & 0.239 & 3.205 & 92.7 & 10 & 0.051 \\
		12 & 4 & 3.585 & 3.277 & 91.4 & 8 & 0.204 & 3.377 & 94.2 & 10 & 0.050 \\
		13 & 4 & 3.700 & 3.444 & 93.1 & 8 & 0.118 & 3.516 & 95.0 & 9 & 0.028 \\
		14 & 4 & 3.807 & 3.635 & 95.5 & 6 & 0.076 & 3.672 & 96.4 & 8 & 0.032 \\
		15 & 4 & 3.907 & 3.814 & 97.6 & 2 & 0.025 & 3.832 & 98.1 & 1 & 0 \\
		16 & 4 & 4 & 4 & 100 & 1 & 0 & 4 & 100 & 1 & 0 \\
		\bottomrule
	\end{tblr}
	\caption{Entanglement generation with the phase set~\eqref{eq:phaseset_pt}, with Bell or cluster states as a resource. $\langle{E}\rangle$ is the expected entanglement, \#$(E)$ is the number of different $E$ occuring in the set, and $\sigma({E})$ is the standard deviation. The values of $\frac{\langle{E}\rangle}{{E}_{d}}$ are given in \%. Decimal numbers are rounded to a fixed number of visible digits. Note that the number of elements (different measurement results) is $2^{2\nu_{\text{max}}}$ in general, or $2^{\nu_{\text{max}}}$ for Bell states by combining equivalent measurement results. The parameter $\xi$ is set to an integer multiple of $2^{\nu_{\text{max}}-1}$ for $\ket{\Phi^{\pm}}$ and $\ket{\Phi_{\cal C}}$, or arbitrary for $\ket{\Psi^{\pm}}$.}
	\label{tab:scheme1}
\end{table}

\begin{figure}[t]
	\centering
	\includegraphics[width=.9\columnwidth]{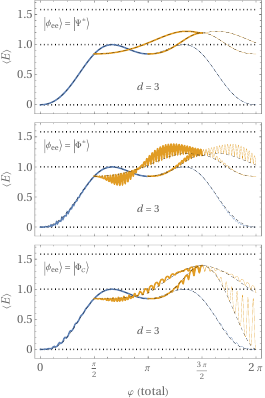}
	\caption{Expected entanglement without postselection for ${d}=3$, for three different types of electron resource states. 
	The electron qubit splitting parameter is chosen to be $\xi=20$ for the solid lines, while the black thin dashed lines are for $\xi=0$.
	}
	\label{fig:Eav3d}
\end{figure}

For completeness, let us first discuss the optimization in the case of Bell states. 
The upper panel of \fref{fig:Eav3d} shows the expected entanglement for $\ket{\phi_{ee}}=\ket{\Psi^{\pm}}$, again for $d=3$ and the phase set \eqref{eq:phaseset_pt}. The phase set~\eqref{eq:phaseset_pt} leads to $\langle{E}\rangle\approx 1.195$; numerical optimization yields $\langle{E}\rangle\approx 1.224$ for the phase set $\{\varphi^{(\nu)}\}\approx(2.673, 1.617)$, i.e., just a slight improvement. 
For all other source states, the $\xi$-dependence of the expected entanglement could generally make the optimization more complicated. The thin dashed lines in \fref{fig:Eav3d} are for the hypothetical case $\xi=0$.
In the case $\ket{\phi_{ee}}=\ket{\Phi^{\pm}}$ the oscillations may indeed enhance $\langle{E}\rangle$ over the $\xi=0$ case, but the optimized~$\langle{E}\rangle$ remains lower than that obtained by $\ket{\phi_{ee}} = \ket{\Phi_{\cal C}}$, so we do not analyze it further.
 
For $\ket{\phi_{ee}} = \ket{\Phi_{\cal C}}$ the case $\xi=0$ is relevant again, because for most $d$ it constitutes an upper bound for the optimized $\langle{E}\rangle$, which is also reached by fine tuning the actual parameter $\xi\gg 1$ to an integer multiple of $2^{\nu_{\text{max}}-1}$. 
We now analyze the question whether $\langle{E}\rangle$ can be optimized, depending on ${d}$. 
For ${d}=2^n$ the scheme~\eqref{eq:phaseset_pt} is perfect, and obviously no optimization is possible. 
For any ${d}=2^n-1$, such as $d=3$ and $d=7$, we also find that no optimization is possible; further, for this phase set all measurement outcome series lead to the same $E$, see Table~\ref{tab:scheme1}.
For other ${d}$ we find that just marginal optimization is possible and requires just tiny corrections applied to one or several phases $\varphi^{(\nu)}$ from set~\eqref{eq:phaseset_pt}; these corrections also raise the number of different values $E$ in the outcome set. 
Furthermore, these corrections also break the above-mentioned periodicity of $\langle{E}\rangle$ in~$\xi$, so that many parameters affect the optimization procedure; however, it is roughly sufficient to consider the case $\xi=0$ because the corrections are so tiny that the periodicity still holds approximately, as becomes evident from the following examples.
For $d=5$ and $\xi=0$, two out of three phases are slightly altered, $\{\varphi^{(\nu)}\}=(\pi, \frac{\pi}{2}+0.091, \frac{\pi}{4}-0.033)$, leading to $\langle{E}\rangle\approx 2.072$, which corresponds to an improvement of $\sim 0.3\%$ over the result $2.065$ from Table~\ref{tab:scheme1}.
For $d=5$, but with $\xi=20$, the same two phases are slightly altered, $\{\varphi^{(\nu)}\}=(\pi, \frac{\pi}{2}+0.028, \frac{\pi}{4}-0.008)$, leading to $\langle{E}\rangle\approx 2.073$, so the difference to $\xi=0$ is negligible.
For $d=6$ and $\xi=0$, the optimal phases are $\{\varphi^{(\nu)}\}=(\pi, \frac{\pi}{2}-0.020, \frac{\pi}{4}-0.0006)$, leading to an improvement of only $\sim 0.02\%$ over the result $\langle{E}\rangle\approx 2.372$.
Therefore, we conclude at least for up to $d=8$ that the phase set~\eqref{eq:phaseset_pt} is exactly equal or very close to the optimal one for maximizing $\langle{E}\rangle$ without postselection, if we set~$\xi$ to any one of the (nearly) optimal values at integer multiples of $2^{\nu_{\text{max}}-1}$.

\section{Multipartite entanglement}
\label{sec:multipartite}

The extension of our scheme to more than two nodes works to some extent analogously to the case $d=2$ described for cluster states in Ref.~\cite{Nemoto2014}.
This means that genuine multipartite entanglement may be generated by successively establishing entanglement links between each two linked nodes of a cluster, where we denote the number of nodes by $M$.

As a simple example, consider a row of $M=3$ nodes labeled $(a),(b),(c)$, the case of pairwise Bell states used as the resource, $\ket{\phi_{ee}} = \ket{\Psi^{+}}$, and the special case of postselection $j_{K}^{(\nu)}=j_{K'}^{(\nu)}$ for any $\nu$ and the two nodes ${K},{K'}$ being entangled. 
With nuclear spins initialized in $\ket{+_{d}}$, a two-qudit Bell state on nodes $(a)$ and $(b)$ can be generated with $P=1/{d}$ for any ${d}$ using phase set~\eqref{eq:phaseset_pt}, i.e., by the operation
\begin{equation}
	{\cal Q}^{(a,b)} = \prod_{\nu=1}^{\lceil E_{d} \rceil}{\cal T}_{j_{a}^{(\nu)}=j_{b}^{(\nu)}}(\ket{\Psi^{+}}\!,2\pi/2^{\nu}).
\end{equation}
Then the successive application of this scheme to pairs $(a,b)$ and $(b,c)$ corresponds to
\begin{equation}
	\left(\mathds{1}_{d}^{(a)}\otimes{\cal Q}^{(b,c)}\right)\cdot
	\left({\cal Q}^{(a,b)}\otimes\mathds{1}_{d}^{(c)}\right)
\end{equation}
and results in a tripartite qudit Greenberger-Horne–Zeilinger (GHZ) state,
$(1/\sqrt{d}) \sum_{i=0}^{{d}-1}\ket{i}\otimes\ket{i}\otimes\ket{i}$, with success rate~$1/{d}^2$.
We note that since ${\cal Q}$ is diagonal, the order of building up entanglement does not matter; for example, we could first generate 1~ebit for each pair $(a,b)$ and $(b,c)$, and then add further entanglement to both pairs. 

The same principle can be applied to longer chains of modules (higher $M$), if each two neighbouring sites can be addressed with this scheme. The special case ($\ket{\Psi^{+}}$, ${j_{K}^{(\nu)}=j_{K'}^{(\nu)}}$) above then results in an $M$-partite qudit GHZ state, which could find application in several quantum networking proposals in QC. 

For $d=2^n$ qudits, the non-postselected outcomes (${j_{K}^{(\nu)}\neq j_{K'}^{(\nu)}}$ for some $\nu,K,K'$) for $M=2$ also result in a maximally entangled state, see Sec.~\ref{sec:scheme_deterministic}. 
Thus, a natural question, but beyond the scope of this paper, is whether the unselected multipartite states ($M>2$) for $d=2^n$ belong to the same entanglement class as the qudit GHZ state. 

\section{Discussion}
\label{sec:discussion}

The nuclear spin quantum memory and transfer scheme presented here may be understood as a higher-dimensional version of the qubit scheme for \isotope{N}{15}V centers~\cite{Everitt2014,Nemoto2014}.
Since our model focusses on the intranode dynamics, a complete benchmark analysis of the architecture is difficult, particularly without specifying the defect center, host material etc.
However, it is worth summarizing the common error sources. 
One error source explicitly included in the model is the transverse hyperfine coupling $A_\perp$. In lowest order, this component results in rather fatal electron-nuclear flip-flop processes, so it needs to be  suppressed by a sufficiently large magnetic field or inversion-symmetry-protected spin-orbital locking (GeV-type systems with low strain), or both. The next-higher-order correction consists of dispersive level shifts, which have been shown to cause a small error, presumably much smaller than other error sources. 

In Sec.~\ref{sec:decoupling} we have qualitatively discussed the errors accumulating on nuclear spins during electron-electron-entangling attempts. 
We still need to discuss the nuclear decoherence due to the coupling with the electron spin; this channel is active during the CPHASE gates, where spin-echo pulses are not allowed.
The $T_2^*$ of the electron spin during this gate (taking 165 ns for \isotope{N}{14}V) is considered the limiting coherence parameter in Appendix~C of Ref.~\cite{Nemoto2014}.
However, this issue does not become more severe for higher nuclear spins. 
The total CPHASE angle from all iterations can be chosen to be smaller than $2\pi$, so it is not much larger than the angle $\pi$ used for $d=2$. Hence, especially with strong coupling, the process is very fast. As an example, $2\pi \hbar/|A_\parallel| \approx 27\, {\rm ns}$ for \isotope{Ge}{73} with $|A_\parallel| \approx 37\, {\rm MHz}$~\cite{Adambukulam2024}, which is much faster than electron coherence times, typically in the range of 10-100~\textmu s at low temperatures.
For the \isotope{N}{14}V center with $I=1$, due to comparable parameter values with the \isotope{N}{15}V center, the error analysis of Ref.~\cite{Nemoto2014} should be qualitatively adoptable.

The strong coupling of truly high nuclear spins can further be leveraged for operating on nuclear spins via hyperfine-selective microwave pulses. This could possibly speed up the single-qudit nuclear gates required for initialization, intermittently (in schemes with driving), or final state correction (prior to entanglement consumption).
In the very strong coupling regime (e.g., 1.475 GHz for \isotope{Bi}{209}), advanced techniques such as optimal control are indispensable to individually address the electron qubit. 

For the diamond defects NV and GeV, the available spin-photon interface can be used for electron-spin initialization, electron-electron entanglement generation, and electron-spin readout~\cite{Nemoto2014}. 
The performance of this interface, in terms of the spin contrast as well as absorption probability, then determines a number of further error sources which are expected to dominate over the errors related with the nuclear memory itself. 
For example, the entangled resource states $\ket{\phi_{ee}}$ could have a low fidelity, and the high-fidelity non-demolition readout via optical photons could require many attempts, resulting in a significant chance of exciting the NV center. 
We do not discuss these errors in more detail as they are not specific to high-spin nuclei; in particular, our (deterministic) scheme does not require more resource states or more readouts per generated distant entanglement than the qubit scheme. 

A feature adopted from Ref.~\cite{Nemoto2014} is the modular structure of the approach, which makes it suitable for the creation of larger clusters. 
We explicitly confirm that genuine multipartite entanglement may be generated successively by applying the scheme to each pair of  neighboring network nodes. 
Apparently, the higher ${d}$, the more rapid is the scaling of the entanglement with the number of parties. 
For the particular application of our scheme in MBQC, the possibility of preparing dedicated states such as higher-dimensional cluster states remains to be investigated.

\section{Summary}
\label{sec:conclusion}

We have analyzed methods to accumulate entanglement in nuclear spin memory qudits by repeated entanglement transfer from their respective electron spin communication qubit, mediated by the native Ising coupling. Based on a universal model, our study is applicable to a wide range of platforms, such as the \isotope{Ge}{73}V center in diamond or the \isotope{Sb}{123} donor in silicon. The protocol that we consider to have the highest practical relevance  is described by a universal phase set, \eref{eq:phaseset_dt}, and generates maximal entanglement deterministically for $d=2^n$ qudits. 
The application of our scheme to larger quantum networks opens the possibility of realizing higher-dimensional qudit versions of several QC protocols.
We have also pointed out some crucial properties of entanglement transfer which could be of importance even in the case~$d=2$.

\section*{Acknowledgements}
 We acknowledge funding from the German Federal Ministry of Education and Research (BMBF) under Grant Agreement No.~13N16212 (SPINNING), and the European Union under
grant agreement No.~101186889 (QuSPARC).

\appendix

\section{Conditions for deterministic generation of maximal entanglement}
\label{sec:criteria_derivation_max}

Here we evaluate a necessary and sufficient condition for generating ${d}$-ranked maximal entanglement in some iteration~$\nu$ (the final one if successful) of entanglement transfer.
If $\ket{\psi^{(\nu)}_{j_a j_b}}$ is rank-${d}$ maximally entangled, this means that the matrix $({{d}/P^{(\nu)}_{j_a j_b}})^{1/2}\tilde{\psi}^{(\nu)}_{j_a j_b}$ is unitary. The coefficient matrix~$\tilde{\psi}^{(\nu)}_{j_a j_b}$ is given for a general two-qubit resource state~$\ket{\phi_{ee}^{(\nu)}}$ in \eref{eq:psinu_S}.
Evaluating the equation 
$\tilde{\psi}^{(\nu)}_{j_a j_b} \big(\tilde{\psi}^{(\nu)}_{j_a j_b}\big)\hc = (P^{(\nu)}_{j_a j_b}/d) \mathds{1}_{d}$,
we obtain the following condition ,
\begin{align}
	\Lambda_{K,00} - & (-1)^{j_a} \Lambda_{K,10} - (-1)^{j_b} \Lambda_{K,01} +(-1)^{j_a+j_b} \Lambda_{K,11} \nonumber \\
	&= \frac{4}{d} P^{(\nu)}_{j_a j_b} \mathds{1}_{d},
	\label{eq:cond}
\end{align}
for $K=a$,
where the matrices are defined as
\begin{subequations}
	\label{eq:cterms}
	\begin{align}
		\Lambda_{a,00} &\equiv (c_{00}^2+c_{01}^2) S_\downarrow  + (c_{10}^2+c_{11}^2) {\cal U}_aS_\downarrow {\cal U}_a\hc , \\
		\Lambda_{a,10} &\equiv (c_{00}c_{10}+c_{01}c_{11}) S_\downarrow {\cal U}_a\hc + \text{h.c.} , \\
		\Lambda_{a,01} &\equiv c_{00}c_{01} \sqrt{S_\downarrow} {\cal U}_b^* \sqrt{S_\downarrow}  + c_{10}c_{11} {\cal U}_a\sqrt{S_\downarrow} {\cal U}_b^* \sqrt{S_\downarrow} {\cal U}_a\hc \nonumber\\
		&\qquad\qquad\qquad\qquad\qquad\qquad\qquad\ + \text{h.c.} , \\
		\Lambda_{a,11} &\equiv c_{00}c_{11} \sqrt{S_\downarrow} {\cal U}_b^* \sqrt{S_\downarrow} {\cal U}_a\hc + c_{01}c_{10} \sqrt{S_\downarrow}{\cal U}_b\tran  \sqrt{S_\downarrow} {\cal U}_a\hc  \nonumber\\
		&\qquad\qquad\qquad\qquad\qquad\qquad\qquad\ + \text{h.c.} .
	\end{align} 
\end{subequations}
Without loss of generality we have taken the coefficients $c_{j_a j_b}$ of the two-qubit state, \eref{eq:2qubitstate}, to be real.
We do not make any assumption about its entanglement $E_{ee}$.
Similarly, from 
$\big(\tilde{\psi}^{(\nu)}_{j_a j_b}\big)\hc \tilde{\psi}^{(\nu)}_{j_a j_b} = (P^{(\nu)}_{j_a j_b}/d) \mathds{1}_{d}$
we obtain \eref{eq:cond} for $K=b$, where the matrices $\Lambda_{b,jj'}$ for node $b$ are obtained from the matrices $\Lambda_{a,jj'}$ by swapping nodes ($a\leftrightarrow b$) and coefficient indices ($c_{jj'}\leftrightarrow c_{j'j}$).

We investigate the condition for deterministic entanglement generation, in which case \eref{eq:cond} needs to hold for all four combinations of $(j_a,j_b)$. This allows us to solve the four equations for the four matrices (each for $K=a,b$) to obtain that each one is the identity matrix multiplied with a scalar, namely,
\begin{subequations}
	\label{eq:cond_gen1}
	\begin{align}
		\label{eq:cond_gen1_a}
		\Lambda_{K,00} &= \frac{1}{d} \mathds{1}_{d}, \\
		\label{eq:cond_gen1_b}
		\Lambda_{K,10} &= -\frac{2}{d} \left(P^{(\nu)}_{00}+P^{(\nu)}_{01}-P^{(\nu)}_{10}-P^{(\nu)}_{11}\right) \mathds{1}_{d} , \\
		\label{eq:cond_gen1_c}
		\Lambda_{K,01} &= \frac{2}{d} \left(P^{(\nu)}_{01}-P^{(\nu)}_{10}\right) \mathds{1}_{d}, \\
		\label{eq:cond_gen1_d}
		\Lambda_{K,11} &= \frac{2}{d} \left(P^{(\nu)}_{00}+P^{(\nu)}_{11}-\frac{1}{2}\right) \mathds{1}_{d}.
	\end{align} 
\end{subequations}
We start by inspecting the first equation, \eqref{eq:cond_gen1_a}. Defining $c_{00}^2+c_{01}^2=1-c_{10}^2+c_{11}^2\equiv\tau_a/2$ and $c_{00}^2+c_{10}^2=1-c_{01}^2+c_{11}^2\equiv\tau_b/2$ (with $0\le \tau_K\le 2$) we can express it as
\begin{equation}
	\label{eq:cond_main_tmp}
	{\cal U}_K\hc \tau_K S_\downarrow {\cal U}_K = \frac{2}{d} \mathds{1}_{d} - (2-\tau_K) S_\downarrow.
\end{equation}
This equation puts constraints on the Schmidt coefficients delivered by the previous iteration $\nu-1$. 
Since ${\cal U}_K$ is unitary, the entries on the diagonal of $\tau_K S_\downarrow$ must be, up to reordering, the same as those on the r.h.s.~of the equation. However, the diagonal entries on the r.h.s.~are reversely ordered in comparison with that of $\tau_K S_\downarrow$, and it follows that $\tau_K S_\downarrow+(2-\tau_K) S_\uparrow = (2/{d})\mathds{1}_{d}$, where $S_\uparrow$ is obtained from $S_\downarrow$ by reversing the order of diagonal elements. 
Thus, the diagonal entries of $S_\downarrow$ (and $S_\uparrow$) come in pairs $\{1/{d} + (2-\tau_K) \varepsilon_i, 1/{d} - \tau_K \varepsilon_i\}$ (with $0\le \varepsilon_i $) %
for $i=1,\dots,\lfloor {d}/2 \rfloor$, and one further central element~$1/{d}$ %
in the case of odd ${d}$. 
Now, since the diagonal elements are squared Schmidt coefficients or zeros, they must sum up to 1, which yields $\tau_a = \tau_b =1$.
Therefore, the paired diagonal elements of $S_\downarrow$ are $1/{d} \pm \varepsilon_i$ with $0\le \varepsilon_i \le 1/{d}$, for $i=1,\dots,\lfloor {d}/2 \rfloor$, and one further unpaired central element $1/{d}$ %
in the case of odd ${d}$.
(This is not a constraint for $d=2$.)
The condition \eqref{eq:cond_main_tmp} can now be restated as 
\begin{equation}
	\label{eq:cond_main}
	{\cal U}_K\hc S_\downarrow {\cal U}_K = \frac{2}{d} \mathds{1}_{d} - S_\downarrow = S_\uparrow
\end{equation}
In the following, the term block refers to a `degenerate' subspace in which diagonal elements of $S_\downarrow$ (or $S_\uparrow$) are identical. We take $E^{(\nu-1)}<E_{d}$ as we want to increase the qudit entanglement, so there is at least one non-vanishing $\varepsilon_i$ and hence there are at least two blocks. 
\eref{eq:cond_main} implies that the matrices ${\cal U}_K$ must be block-antidiagonal, each block being unitary. The swapped version of this equation, ${\cal U}_K\hc S_\uparrow {\cal U}_K = S_\downarrow$, must also hold since blocks are arranged symmetrically with respect to the center.

Another constraint derives on the entanglement resource state $\ket{\phi_{ee}}$.
Since $\tau_a = \tau_b =1$, it must have the form
\begin{equation}
	\label{eq:cond_phi_ee}
	\phi_{ee} = \begin{pmatrix}
		\sigma_{00}{c} & \sigma_{01}\sqrt{1/2-{c}^2} \\ 
		\sigma_{10}\sqrt{1/2-{c}^2} & \sigma_{11}{c}
	\end{pmatrix}
\end{equation}
with $0\le {c} \le 1/\sqrt{2}$ and arbitrary sign factors $\sigma_{j_a j_b}$.
Let us define $\kappa_\sigma\equiv(\sigma_{00}\sigma_{11}+\sigma_{01}\sigma_{10})/2\in \{-1,0,1\}$ and $\kappa_c\equiv {c} \sqrt{1/2-{c}^2}$.
An EPR pair (any maximally entangled two-qubit state, taking coefficients real) is of the form \eqref{eq:cond_phi_ee} as its matrix $\sqrt{2}\phi_{ee}$ is unitary, and is characterized by $\kappa_\sigma\kappa_c=0$; in particular $\kappa_c=0$ corresponds to a Bell state and $\kappa_\sigma=0$ together with $\kappa_c=1/4$ to a two-qubit cluster state.

Eqs.~\eqref{eq:cond_gen1_b}-\eqref{eq:cond_gen1_d} can now be written as
\begin{widetext}
\begin{subequations}
	\label{eq:cond_gen2}
	\begin{align}
		\label{eq:cond_gen2_a}
		-\sigma_{10}\sigma_{11} \kappa_\sigma\kappa_c ({\cal U}_K S_\downarrow + \text{h.c.}) &= \frac{2}{d} \left(P^{(\nu)}_{00}+P^{(\nu)}_{01}-P^{(\nu)}_{10}-P^{(\nu)}_{11}\right) \mathds{1}_{d} , \\
		\label{eq:cond_gen2_b}
		\sigma_{00}\sigma_{01}\kappa_c \sqrt{S_\downarrow}{\cal U}_b^* \sqrt{S_\downarrow} + \sigma_{10}\sigma_{11}\kappa_c {\cal U}_a\sqrt{S_\downarrow}{\cal U}_b^* \sqrt{S_\downarrow} {\cal U}_a\hc + \text{h.c.} &= \frac{2}{d} \left(P^{(\nu)}_{01}-P^{(\nu)}_{10}\right) \mathds{1}_{d}, \\
		\label{eq:cond_gen2_c}
		\sigma_{00}\sigma_{11} c^2 \sqrt{S_\downarrow}{\cal U}_b^* \sqrt{S_\downarrow} {\cal U}_a\hc + \sigma_{01}\sigma_{10} (\tfrac{1}{2}-c^2) \sqrt{S_\downarrow}{\cal U}_b\tran \sqrt{S_\downarrow} {\cal U}_a\hc + \text{h.c.} &= \frac{2}{d} \left(P^{(\nu)}_{00}+P^{(\nu)}_{11}-\frac{1}{2}\right) \mathds{1}_{d}.
	\end{align} 
\end{subequations}
\end{widetext}
These equations must also hold when swapping $a\leftrightarrow b$ and $\sigma_{01}\leftrightarrow\sigma_{10}$.
The $S$-matrices may be commuted to the left using the generalized commutation relation \eqref{eq:cond_main} along with its swapped version ($S_\downarrow \leftrightarrow S_\uparrow$).
On the l.h.s.~of Eqs.~\eqref{eq:cond_gen2_a},\eqref{eq:cond_gen2_b}, each term contains a product of an odd number of ${\cal U}_K$-matrices and is thus block-antidiagonal (in particular, since there are at least two blocks, the upper left element of the matrix is zero), while the r.h.s.~is diagonal, hence both sides must vanish. From both r.h.s.~we obtain that $P^{(\nu)}_{00}=P^{(\nu)}_{11}\equiv P^{(\nu)}_\text{eq}/2$ and $P^{(\nu)}_{01}=P^{(\nu)}_{10} = (1-P^{(\nu)}_\text{eq})/2$.
Note that this is not a separate constraint, i.e., it is fulfilled for any state $\ket{\psi^{(\nu-1)}}$ and conditional operations $U_K$.
Setting the l.h.s.~of \eref{eq:cond_gen2_a} equal to zero, we obtain that either $\ket{\phi_{ee}}$ is an EPR pair ($\kappa_\sigma\kappa_c=0$), or ${\cal U}_K S_\downarrow + {\cal U}_K\hc S_\uparrow=0$. There is no unitary ${\cal U}_K$ that fulfills the latter equation unless all $\varepsilon_i$ vanish, which we have excluded. Hence, $\ket{\phi_{ee}}$ must be an EPR pair. This is trivial in the case $E^{(\nu-1)}=E^{(\nu)}-1$ (with $E^{(\nu)} = E_{d}$), %
but not for higher $E^{(\nu-1)}$. 
We are now left with
\begin{subequations}
	\label{eq:cond_gen3}
	\begin{align}
		\label{eq:cond_gen3_a}
		\kappa_c \sqrt{S_\downarrow S_\uparrow} \left[ {\cal U}_a, \left({\cal U}_b + {\cal U}_b\hc\right)\tran \right] &= 0, \\
		\label{eq:cond_gen3_b}
		\sqrt{S_\downarrow S_\uparrow} \left[c^2 {\cal U}_a {\cal U}_b\tran - (\tfrac{1}{2}-c^2) {\cal U}_a {\cal U}_b^*  + \text{h.c.} \right] &= \bar{\sigma}_c \frac{2P^{(\nu)}_\text{eq}-1}{d} \mathds{1}_{d},
	\end{align} 
\end{subequations}
where $\bar{\sigma}_c \equiv\sigma_{00}\sigma_{11}$ for $c\neq 0$, and $\bar{\sigma}_c \equiv-\sigma_{01}\sigma_{10}$ for $c=0$ ($\ket{\phi_{ee}}=\ket{\Psi^{\pm}}$); 
we have used that $\sigma_{00}\sigma_{01}\sigma_{10}\sigma_{11}=-1$, which follows from $\kappa_\sigma=0$ or can be freely chosen in the case $\kappa_c=0$.
Eqs.~\eqref{eq:cond_gen3} must also hold when replacing $a\leftrightarrow b$. 

In the case $E^{(\nu-1)}=E^{(\nu)}-1$, the goal considered here is equivalent with that from Sec.~\ref{sec:condition_transfer_deterministic}, with $E_{ee}=1$. Indeed, under the constraints derived from \eref{eq:cond_main_tmp}, this case requires that ${d}$ is even and $\ket{\psi^{(\nu-1)}}$ is $r^{(\nu-1)}={d}/2$-ranked maximally entangled,
hence $\varepsilon_i = 1/{d}$ for all $i$, and the matrix $S_\downarrow S_\uparrow$ is zero, so 
that Eqs.~\eqref{eq:cond_gen3} are fulfilled 
and the constraint \eqref{eq:cond_main} 
reduces to \eref{eq:cond_complete_calU}.

The other case is $E^{(\nu-1)}>E^{(\nu)}-1$, which means that $r^{(\nu-1)}>{d}/2$ and the available target Hilbert space for entanglement transfer is too small to fit the entanglement of the EPR pair. 
We can further distinguish the case $r^{(\nu-1)}<{d}$, in which the outermost diagonal blocks of $S_\downarrow S_\uparrow$ are zero and hence, from \eref{eq:cond_gen3_b} we obtain $P^{(\nu)}_\text{eq}=1/2$, and the case $r^{(\nu-1)}={d}$, in which $S_\downarrow S_\uparrow$ is invertible. 
In both cases, Eqs.~\eqref{eq:cond_gen3} can be further processed by inspecting diagonal blocks individually. 
We refer to the antidiagonal blocks of ${\cal U}_a$ (${\cal U}_b$) as $\{{\cal A}_{\bar{\varepsilon}}\}$ ($\{{\cal B}_{\bar{\varepsilon}}\}$). Here, $\bar{\varepsilon}= +\varepsilon$ for the blocks in the upper right quarter, $\bar{\varepsilon} = -\varepsilon$ for the ones in the lower left,  and $\bar{\varepsilon} = 0$ for the central block (if existent). 
Note that the central block must exist for odd ${d}$.
For outermost blocks in the case $r^{(\nu-1)}<{d}$ (i.e., blocks with $\pm\varepsilon=\pm 1/{d}$) no constraint (except unitarity) is obtained.
For other blocks (in both cases) we define
$\zeta_{\varepsilon}^{(\nu)} \equiv \bar{\sigma}_c  (P^{(\nu)}_\text{eq}-1/2)/\sqrt{1 - {d}^2\varepsilon^2}$.
It is possible to simplify the constraints given in Eqs.~\eqref{eq:cond_gen3} by specifying the resource state $\ket{\phi_{ee}}$.
For $c=1/\sqrt{2}$ (Bell state, $\ket{\phi_{ee}}=\ket{\Phi^{\pm}}$) we obtain that 
the real part of the eigenvalues of the matrices ${\cal A}_{\bar{\varepsilon}} {\cal B}_{\bar{\varepsilon}}\tran$ must be 
$\zeta_{\varepsilon}^{(\nu)}$;
this means zero in the case $r^{(\nu-1)}<{d}$.
For $c=0$ (Bell state, $\ket{\phi_{ee}}=\ket{\Psi^{\pm}}$) we obtain the same eigenvalue constraint but for the matrices ${\cal A}_{\bar{\varepsilon}} {\cal B}_{-\bar{\varepsilon}}^*$.
For $\kappa_c \neq 0$ (other states $\ket{\phi_{ee}}$) the constraints do not simplify that much. 
One further special case is $c=1/2$ (two-qubit cluster state, $\ket{\phi_{ee}}=\ket{\Phi_{\cal C}}$), for which we obtain that (still for $|\bar{\varepsilon}|<1/d$ only) the real part of the eigenvalues of the matrices ${\cal A}_{\bar{\varepsilon}} ({\cal B}_{\bar{\varepsilon}}-{\cal B}_{-\bar{\varepsilon}}\hc)\tran$ must be 
$2\zeta_{\varepsilon}^{(\nu)}$,
as well as the condition
\begin{equation}
	\label{eq:cond_cluster_a}
	{\cal A}_{\bar{\varepsilon}} ({\cal B}_{\bar{\varepsilon}} + {\cal B}_{-\bar{\varepsilon}}\hc)\tran - ({\cal B}_{-\bar{\varepsilon}} + {\cal B}_{\bar{\varepsilon}}\hc)\tran {\cal A}_{-\bar{\varepsilon}} = 0.
\end{equation}
Again, the given conditions for each case are necessary and sufficient for generating ${d}$-ranked maximal entanglement starting from $E^{(\nu-1)}>E^{(\nu)}-1$.

\section{Defect center models}
\label{sec:defectmodels}

In this Section we derive effective Hamiltonians for two defect centers: the GeV center in diamond and the vanadium defect in silicon carbide.

\subsection{Germanium vacancy center in diamond}
\label{sec:GeV}

The GeV is a split vacancy with an orbital degree of freedom. 
Following Ref.~\cite{Adambukulam2024}, we derive a low-energy effective Hamiltonian.
The magnetic field ${B}$ is assumed to be aligned along the $z$-axis.
The isolated electron Hamiltonian in the basis $\{\ket{e_{+}, \uparrow}, \ket{e_{-}, \downarrow}, \ket{e_{-}, \uparrow}, \ket{e_{+}, \downarrow}\}$ is
\begin{equation}
	H_{e} = \frac{1}{2} 
	\begin{pmatrix} 
	\lambda + \gamma_e {B} & 0 & 2\epsilon & 0  \\
	0 & \lambda - \gamma_e {B} & 0 & 2\epsilon^* \\
	2\epsilon^* & 0 & -\lambda + \gamma_e {B} & 0 \\
	0 & 2\epsilon & 0 & -\lambda - \gamma_e {B}  \\
	\end{pmatrix} 
\end{equation}  
where $\lambda\sim$ 165~GHz~\cite{Maity2018} is spin-orbit coupling and $\epsilon = \alpha - i\beta$ is a strain parameter. 
The orbital Zeeman effect is neglected as it is quenched.
The hyperfine Hamiltonian $H_{\rm{hf}}$ is given by \eref{eq:H_hf}, with $I=9/2$ for \isotope{Ge}{73}. 
The isolated nuclear spin Hamiltonian $H_{n}$, see \eref{eq:H_nuc}, is not required for the reasons outlined in Sec.~\ref{sec:node_model}. 

For $\gamma_e {B}, |\epsilon| \ll \lambda$ there are two blocks 
$\{\ket{e_{+}, \uparrow}, \ket{e_{-}, \downarrow}\}$
and $\{\ket{e_{-}, \uparrow}, \ket{e_{+}, \downarrow}\}$,
which are energetically far separate from each other. 
The two blocks are coupled by the strain~$\epsilon$ and the perpendicular hyperfine component $A_\perp$. 
Treating the coupling by a Schrieffer-Wolff transformation, we obtain an effective Hamiltonian $H^{\rm{eff}}_{}=H^{\rm{eff}}_{e}+H^{\rm{eff}}_{\rm{hf}}$ for the lower energy block
\begin{align}	
	H^{\rm{eff}}_{e} &= 
	-\frac{\lambda}{2}
	+\gamma_e {B} S_z
	-4|\epsilon|^2, \\
	H^{\rm{eff}}_{\rm{hf}} &= 
	A_\parallel S_z \otimes I_z -
	\frac{1}{\lambda}
	\begin{pmatrix} 
	0 & \epsilon^* A_\perp I_- \\
	\epsilon A_\perp I_+ & 0
	\end{pmatrix}.
\end{align}
Removing the constant (spin-independent) terms, we can write
\begin{align}
	H^{\rm{eff}} &= \gamma_e {B} S_z + \mathbf{S} \mathbf{A} \mathbf{I},
\end{align}
with the hyperfine tensor
\begin{align}
	\mathbf{A} &= 
			\begin{pmatrix}
				-\alpha A_\perp/\lambda & -\beta A_\perp/\lambda & 0 \\
				\beta A_\perp/\lambda & -\alpha A_\perp/\lambda & 0 \\
				0 & 0 & A_\parallel
			\end{pmatrix}.
\end{align}
To remove the perpendicular hyperfine component, we consider the parameter regime
$(|\epsilon| A_\perp/\lambda)^2 \ll |\gamma_e {B} A_\parallel|$, which seems realistic.
This allows us to perform another Schrieffer-Wolff transformation.
In zeroth order, we are left with the Hamiltonian,
\begin{align}
	\label{eq:Heff_GeV}
	H^{\rm{eff}} &= \gamma_e {B} S_z + A_\parallel S_z I_z.
\end{align}
For vanishing strain $|\epsilon|$ this result is exact, while otherwise, higher-order corrections appear.
These corrections have only a marginal influence on the generated entanglement, which can be seen from an evaluation analogous to that in App.~\ref{sec:hf_perp} for the NV-type center.

The Hamiltonian \eqref{eq:Heff_GeV} is of the desired form, i.e., it is consistent with the model from Sec.~\ref{sec:node_model}. The Zeeman term may be removed by transforming to a rotating frame and the term $A_\parallel S_z I_z$ allows us to implement a CPHASE-like gate.

\subsection{Vanadium defect in silicon carbide}
\label{sec:Va_in_SiC}
Another qudit memory is provided by the V defect in SiC, with nuclear spin $I=7/2$ for the isotope $^{51}V$ \cite{Tissot2021a,Tissot2021b}. The level structure of the Kramers doublets formed by the combination of the crystal field, spin-orbit coupling, and strain has been studied in detail in Ref.~\cite{Tissot2024}. The resulting effective Hamiltonian 
$H^{\rm{eff}}_{}=H^{\rm{eff}}_{e}+H^{\rm{eff}}_{\rm{hf}}$
is, in this case,
\begin{align}	
	H^{\rm{eff}}_{e} &= 
	\gamma_e {B} S_z, \label{eq:Ve}\\
	H^{\rm{eff}}_{\rm{hf}} &= 
	\mathbf{S} \mathbf{A} \mathbf{I} ,\label{eq:Vhf}
\end{align}
where an irrelevant additive constant in \eqref{eq:Ve} was omitted, and where the hyperfine tensor $\mathbf{A}$, assuming strain in $x$ direction, is given by Eq.~(23) of Ref.~\cite{Tissot2024},
\begin{align}	
	\mathbf{A}= \begin{pmatrix}
    a^{xx}+a^{xy} & 0 & a^{xz} \\
    0 & a^{xx}-a^{xy} & 0 \\
    -a^{zx} & 0 & a^{zz}
    \end{pmatrix}.
\end{align}
For the ground-state doublet $|1,-,\sigma\rangle$, 
one finds for the matrix elements 
$a^{ij}=a^{ij}_{1,-}$ the following expressions,
$a^{zz}=a_{11}^z-2{a^z_{11}}''\cos(\theta_1)$,
$a^{zx}={a_{11}^x}'\sin(\theta_1)$,
$a^{xy}=-a_{11}^x [1+\cos(\theta_1)]$,
$a^{xz}={a_{11}^x}'[1-\cos(\theta_1)]$,
and $a^{xx}=-{a_{11}^z}'\sin(\theta_1)$ \cite{Tissot2024}.
In the case of strain in the $x$ direction, the strain mixing angle
is given by $\tan(\theta_1)= 2\epsilon_{11}^x/\lambda_{11}^z$
where $\epsilon_{11}^x = s_{11}^x \epsilon_{xz}+{s_{11}^x}'(\epsilon_{yy}-\epsilon_{xx})/2$ denotes the reduced component of the strain 
tensor and $\lambda_{11}^z = 529\,{\rm GHz}$ the relevant spin-orbit coupling constant.
Here, $\epsilon_{ij}$ are the strain tensor matrix elements and $s_{11}=251\,{\rm THz}$ 
and $s_{11}'=230\,{\rm THz}$ are strain-orbital coupling coefficients.

For small strain, $\theta_1/\pi \lesssim 0.1$, and magnetic fields $B\gg 10\,{\rm mT}$,
we are in the regime $\gamma_e B \gg a^{xx}, a^{xy}, a^{zz} \gg a^{xz}, a^{zx}$.
In this case, the hyperfine Hamiltonian acquires a diagonal form similar to \eref{eq:H_hf}, 
with the parallel hyperfine coupling $A_\parallel \approx a^{zz}$.  For unstrained SiC, $a^{zz} = 232\,{\rm MHz}$, while in strained SiC, the
coupling decreases monotonically to $a^{zz} = 201\,{\rm MHz}$ for $\theta_1=\pi/2$ \cite{Tissot2024}.
However, the transverse diagonal components are not renormalized by a ratio of strain and spin-orbit coupling as for the GeV. 
The isolated consideration of the ground state Kramers doublet restricts the Zeeman energy $\gamma_e {B}$ to values much smaller than the energy of the first excited Kramers doublet, which is in the range of 10-500~GHz, depending on the SiC polytype and defect site \cite{Wolfowicz2020}. Therefore, at least for the $\alpha$-site V defect, the suppression of $A_\perp$ components works to some degree, even though the higher-order corrections, determined by level shifts on the order $A_\perp^2/A_\parallel/\gamma_e {B}$ as derived in Sec.~\ref{sec:hf_perp}, could become more significant than in the NV (and group-IV vacancy) center model.  

\section{Effect of exchange coupling on nuclear-spin entanglement}
\label{sec:hf_perp}

The transverse hyperfine component $A_\perp$ in our protocol constitutes an undesirable coupling between electron and nuclear spins. 
While its magnitude is similar to the component $A_\parallel$, its effect may be suppressed by tuning into a parameter regime where consecutive electron spin states are split by energies much larger than $|A_\perp|$, so that a Schrieffer-Wolff approximation is appropriate. 
This is carried out in the following for the NV center model described in Sec.~\ref{sec:node_model}, keeping the electron and nuclear spin quantum numbers ($s$ and $I$, respectively) general in the beginning. The transformation works analogously for other defect center models in appropriate parameter regimes.

We first split the Hamiltonian $H$ from \eref{eq:H_tot} into the unperturbed part $H_0$, which is secular and thus block-diagonal (here, even diagonal),
\begin{align}
	{H}_{0} &=  {D} {S}_z^2 + \gamma_e {B} {S}_z + A_\parallel S_z I_z
	+ Q {I}_z^2 - \gamma_n {B} {I}_z,
\end{align}
and the block-off-diagonal perturbation
\begin{align}
	{V} &= \dfrac{A_\perp}{2} (S_- I_+ + S_+ I_-),	
\end{align}
describing flip-flop processes. 
At this point, we do not yet project the Hamiltonian onto the electron spin qubit subspace, because the flip-flop transition to the non-qubit state may in general be just as important as the qubit transition. 

We want an effective Hamiltonian $H^{\rm{eff}}=e^{\cal S}He^{-\cal S}$ that is block-diagonal to first order in $A_\perp$. This is the case if the anti-Hermitian generator ${\cal S}$ satisfies $[H_0,{\cal S}]=V$. 
The solution strongly simplifies if we neglect the contributions of the last three terms in $H_0$ (which are typically smaller than the first two electron-only terms) in the occurring energy denominators, i.e., approximating the zeroth-order energy of the electron spin by~${\epsilon_{m_s} \approx {D} m_s^2 + \gamma_e {B} m_s}$.
Then we find
\begin{equation}
	{\cal S} = \dfrac{A_\perp}{2} S_+ \left(\sum_{m_s=-s}^{s-1}\frac{\ket{m_s}\bra{m_s}}{\varepsilon_{m_s+1}-\varepsilon_{m_s}}\right)I_- - \text{h.c.}
\end{equation}
The effective Hamiltonian is then found in second-order perturbation theory~\cite{Winkler2003} as $H^{\rm{eff}} = H_0 + H^{(2)} + {\cal O}(A_\perp^{3})$ with $H^{(2)}=\frac{1}{2}[{\cal S},V]$.
For the diagonal blocks of $H^{(2)}$ we obtain
\begin{align}
	&\bra{m_s}H^{(2)}\ket{m_s} = \\
	&\qquad\qquad\dfrac{A_\perp^2}{2} 
	\left(\frac{(1-\delta_{m_s,-s})I_-I_+}{\varepsilon_{m_s}-\varepsilon_{m_s-1}}+\frac{(1-\delta_{m_s,s})I_+I_-}{\varepsilon_{m_s}-\varepsilon_{m_s+1}}\right). \nonumber
\end{align}
This describes virtual transitions between successive nuclear spin levels, weighted by the inverse energy cost of the corresponding electron spin flip-flop transition.
In the qubit case $I=1/2$, the error resulting from this second-order correction can be calibrated away, because the r.h.s.~can be described by ($m_s$-dependent) scalar and $I_z$ terms, as those in $H_0$. 
In the case $I>1/2$, the error on a CPHASE gate becomes difficult to correct. However, since the conditional Hamiltonians for the two electron qubit states still commute, the spin-echo technique described in Sec.~\ref{sec:decoupling} also erases an electron-nuclear entangling mechanism resulting from the second-order correction.
Therefore, an error from $A_\perp$ coupling manifests itself mainly during the qubit-qudit entanglement transfer rather than during electron-electron entangling attempts, and this error is estimated below.

To give an explicit example of the Hamiltonian, we consider ${s=1}$ and $I=1$ (\isotope{N}{14}V center) 
and select the states $m_s=0$ and $m_s=+1$ as the qubit subspace. We also take $\xi_a = \xi_b \equiv \xi$ again.
The unperturbed conditional Hamiltonians from \eref{eq:Heff} are
\begin{equation}
	h^{\rm{eff}}_{j} = (-1)^{j} \frac{A_\text{net}}{2} \mathrm{diag}\left( 1+\xi, \xi, -1+\xi \right),
\end{equation}
and the corrections to be added are given by
\begin{align}    
	h^{\rm{eff,corr}}_{1} &= -A_\text{net}\mathrm{diag}\left( 0, \zeta_+, \zeta_+ \right), \\
    h^{\rm{eff,corr}}_{0} &= A_\text{net}\mathrm{diag}\left( \zeta_+, \zeta_++\zeta_-, \zeta_-\right),
\end{align}
with $\zeta_\pm \equiv A_\perp^2 / A_\text{net} / ({D}\pm \gamma_e {B})$. Since we are interested in rough error estimates,  we further  neglect the Zeeman term in the denominator of  $\zeta_\pm$ below, assuming a significantly larger zero-field splitting ${D}$, so we replace $\zeta_\pm$ by $\zeta\equiv A_\perp^2 / A_\text{net} {D}$. 
As an example, for the \isotope{N}{14}V center, $\zeta\approx -1.2 \cdot 10^{-3}$. 
For ${|\zeta|\ll 1}$, the entanglement transferred onto nuclear spin is subject to a small correction of order ${\cal O}(\zeta^2)$, as long as our scheme is sufficiently short, i.e., not extending over many (${\cal O}(1/\zeta)$) periods of $2\pi\hbar/A_\text{net}$. This is obviously the case if we select the phase~$\varphi^{(\nu)}$ for each iteration from within the first period, which also minimizes the impact of decoherence. 

To evaluate the influence of the perpendicular hyperfine component, we pick two relevant cases, for which our scheme in the case $\zeta=0$ results in $E=E_{d}$ without dependence on~$\xi$ (taking $\xi_a = \xi_b \equiv \xi$ again). 
First, we consider the conditional entanglement for $d=3$ and $\ket{\phi_{ee}}=\ket{\Psi^{\pm}}$, obtained by postselection of $j_{a}^{(\nu)}=j_{b}^{(\nu)}$ for all $\nu$. 
For the phases $\varphi^{(1)}=\pi$ and $\varphi^{(2)}=\pi/2$, 
the relative reduction of $E$ from $E_{d}=\log_2{3}$ is about $8.2 \zeta^2$.
This corresponds to a tiny value of about $1.1\cdot 10^{-5}$ for the parameter value of $\zeta$ given above for the \isotope{N}{14}V center. 
The actual maximum of $E$ is shifted to slightly different phases $\{\varphi^{(\nu)}\}$, where the shifts are on the order of~$\zeta$. However, for the optimal phases, $1-E/E_{d}\approx 9.3\cdot 10^{-6}$, so there is hardly any benefit.
Second, we consider the unconditional expected entanglement for $d=4$, where the set of phases $(\pi,\pi/2)$ achieves two consecutive complete transfers. Here, the relative reduction of $E$ from the ideal value $E_{d}=2$ is about $13.8 \zeta^2$. %
These results should be sufficient to estimate the small influence of~$A_\perp$.
For higher qudit dimensions, corrections remain within the same order of magnitude. In a realistic setup, these errors are presumably negligible compared to other error sources not included in our model.

\bibliography{bibl.bib}

\end{document}